\documentclass[english,aps,prl,superscriptaddress,twocolumn]{revtex4}
\usepackage[T1]{fontenc}
\usepackage[latin9]{inputenc}
\usepackage{float}
\usepackage{amsmath}
\usepackage{amssymb}
\usepackage{graphicx}
\usepackage{esint}

\makeatletter

\providecommand{\tabularnewline}{\\}

\@ifundefined{textcolor}{}
{%
 \definecolor{BLACK}{gray}{0}
 \definecolor{WHITE}{gray}{1}
 \definecolor{RED}{rgb}{1,0,0}
 \definecolor{GREEN}{rgb}{0,1,0}
 \definecolor{BLUE}{rgb}{0,0,1}
 \definecolor{CYAN}{cmyk}{1,0,0,0}
 \definecolor{MAGENTA}{cmyk}{0,1,0,0}
 \definecolor{YELLOW}{cmyk}{0,0,1,0}
}

\usepackage{palatino}
\usepackage[hypertexnames=false]{hyperref}

\def\thesection{\Alph{section}}

\AtBeginDocument{\renewcommand{\ref}[1]{\autoref{#1}}}

\makeatother

\usepackage{babel}
\begin{document}

\title{A quantum phase transition from triangular to stripe charge order
in NbSe$_{2}$}

\author{Anjan Soumyanarayanan}

\email{anjan@physics.harvard.edu}

\selectlanguage{english}%

\affiliation{Department of Physics, Harvard University, Cambridge, MA 02138, USA}

\affiliation{Department of Physics, Massachusetts Institute of Technology, Cambridge,
MA 02139, USA}

\author{Michael M. Yee}

\affiliation{Department of Physics, Harvard University, Cambridge, MA 02138, USA}

\author{Yang He}

\affiliation{Department of Physics, Harvard University, Cambridge, MA 02138, USA}

\author{Jasper van Wezel}

\affiliation{Materials Science Division, Argonne National Laboratory, Argonne,
IL 60439, USA}

\affiliation{H. H. Wills Physics Laboratory, University of Bristol, Bristol BS8
1TL, UK}

\author{D.J. Rahn}

\affiliation{Institute of Experimental and Applied Physics, University of Kiel,
24098 Kiel, Germany}

\author{K. Rossnagel}

\affiliation{Institute of Experimental and Applied Physics, University of Kiel,
24098 Kiel, Germany}

\author{E.W. Hudson}

\affiliation{Department of Physics, Pennsylvania State University, State College,
PA 02138, USA}

\author{M.R. Norman}

\affiliation{Materials Science Division, Argonne National Laboratory, Argonne,
IL 60439, USA}

\author{Jennifer E. Hoffman}

\email{jhoffman@physics.harvard.edu}

\selectlanguage{english}%

\affiliation{Department of Physics, Harvard University, Cambridge, MA 02138, USA}
\begin{abstract}
The competition between proximate electronic phases produces a complex
phenomenology in strongly correlated systems. In particular, fluctuations
associated with periodic charge or spin modulations, known as density
waves, may lead to exotic superconductivity in several correlated
materials. However, density waves have been difficult to isolate in
the presence of chemical disorder, and the suspected causal link between
competing density wave orders and high temperature superconductivity
is not understood. Here we use scanning tunneling microscopy to image
a previously unknown unidirectional (stripe) charge density wave (CDW)
smoothly interfacing with the familiar tri-directional (triangular)
CDW on the surface of the stoichiometric superconductor NbSe$_{2}$.
Our low temperature measurements rule out thermal fluctuations, and
point to local strain as the tuning parameter for this quantum phase
transition. We use this discovery to resolve two longstanding debates
about the anomalous spectroscopic gap and the role of Fermi surface
nesting in the CDW phase of NbSe$_{2}$. Our results highlight the
importance of local strain in governing phase transitions and competing
phenomena, and suggest a new direction of inquiry for resolving similarly
longstanding debates in cuprate superconductors and other strongly
correlated materials.

\end{abstract}
\maketitle

\section{Introduction\label{sec:intro}}

While a classical phase transition separates two states of matter
at different temperatures, two ordered ground states of a material
at zero temperature are separated by a quantum critical point (QCP).
The competition between proximate ordered phases near the QCP can
dramatically influence a large region of the phase diagram\cite{Sachdev2000}.
While the fluctuations from competing quantum states lead to exotic
physics even at higher temperatures, low temperature studies of these
states can lead to a better understanding of the root of the competition.
Density waves - charge or spin ordered states of collective origin
driven by instabilities of the Fermi surface (FS) - exist in close
proximity to superconductivity (SC) in several classes of correlated
materials\cite{Norman2005,Johnston2010a,Jerome2002}, and various
proposals have recently emerged to study their interplay in the presence
of strong inhomogeneity in these systems\cite{Kivelson2003}. In this
light, it is surprising that charge density waves (CDWs) are not fully
understood even in the weakly correlated and stoichiometric transition
metal dichalcogenides (TMDCs). While a classic CDW arises from strong
FS nesting, resulting in a sharply peaked susceptibility and a Kohn
anomaly at the CDW wavevector, the quasi-2D TMDCs are known to deviate
from this picture\cite{Johannes2006}.

$2H$-NbSe$_{2}$ is a layered TMDC which has generated much recent
interest\cite{Suderow2005,Kiss2007,Feng2011} as a model system for
understanding the interplay of the CDW and SC phases which onset at
$T_{{\rm CDW}}\sim33{\rm \, K}$ and $T_{{\rm SC}}\simeq7.2{\rm \, K}$
respectively\cite{Moncton1975,Wilson2001}. Despite extensive study\cite{Kiss2007,Borisenko2009,Mialitsin2010,Weber2011,Rahn2012},
several key facts about its familiar tri-directional ($3Q$) CDW remain
unresolved, including the role of FS nesting in determining its wavevector
$\vec{q}_{3Q}$, and the magnitude of the spectral gap and its role
in the energetics of the transition. First, angle-resolved photoemission
(ARPES) studies have been unable to uniquely identify $\vec{q}_{3Q}$-nested
FS regions in $2H$-NbSe$_{2}$\cite{Kiss2007,Borisenko2009,Rahn2012,Straub1999,Rossnagel2001,Valla2004,Shen2008a}.
Meanwhile, recent studies indicate a broadly peaked susceptibility
and a soft phonon over a range of wavevectors around $\vec{q}_{3Q}$\cite{Johannes2006,Weber2011,Rahn2012},
suggesting instead that the $q$-dependence of the electron-phonon
coupling could play an important role in driving the transition. Second,
kinks in tunneling spectra at $\pm35\,{\rm mV}$ ($\pm\varepsilon_{{\rm K}}$),
historically identified as gap edges, correspond to an anomalously
large energy scale for the corresponding $T_{{\rm CDW}}$ $(2\varepsilon_{{\rm K}}/3.5k_{{\rm B}}T_{{\rm CDW}}\sim7.05$)\cite{Hess1991},
while recent ARPES studies indicate a much smaller $\sim3-5\,{\rm mV}$
gap\cite{Borisenko2009,Rahn2012}.

Our discovery, by low-temperature scanning tunneling microscopy (STM),
of a $1Q$ CDW with distinct wavelength and tunneling spectra from
the $3Q$ CDW, in conjunction with band structure calculations, allows
us to resolve both longstanding questions of the wavevector and the
gap. First, the distinct wavelengths demonstrate that FS nesting plays
a negligible role in setting their magnitude. Second, the distinct
tunneling spectra of the $1Q$ CDW region help us disentangle the
$3Q$ CDW spectra to expose a particle-hole asymmetric gap, riding
on top of a strong inelastic background.

\section{Results\label{sec:results}}

\begin{figure*}
\includegraphics[width=6in]{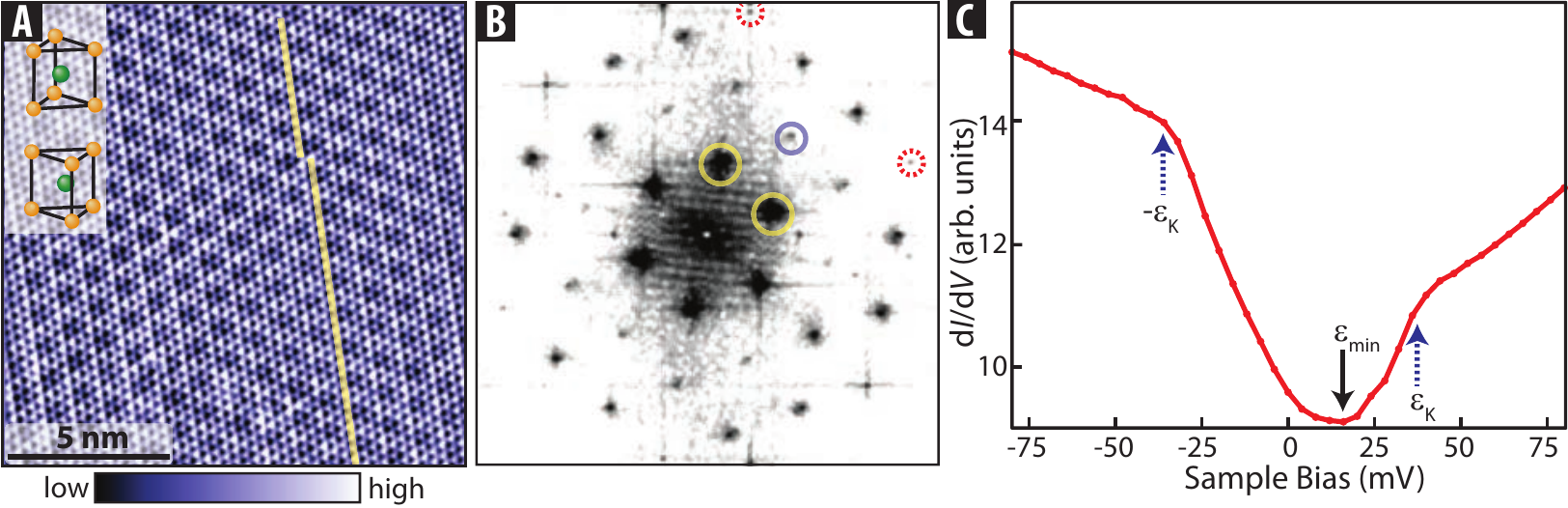}

\caption{\textbf{STM of the 3Q CDW.} (\emph{A}) Atomic resolution topograph
showing the $\sim3\, a_{0}$ periodic CDW. Yellow lines are overlaid
on a one atom shift of the CDW maximum (phase slip). Inset shows the
crystal structure of $2H$-NbSe$_{2}$, with alternating layers of
Nb (green) and Se (orange) atoms. (\emph{B}) Fourier Transform (FT)
of a larger ($\sim45\,{\rm nm}$) topograph, displaying a $3Q$ CDW.
Primary CDW wavevectors, $\vec{{q}_{1}}$ and $\vec{{q}_{2}}$ (yellow
circles), a secondary CDW wavevector, $\vec{{q}_{1}}+\vec{{q}_{2}}$
(blue circle) and Bragg vectors (dashed red circles) are indicated.
(\emph{C}) Average $dI/dV$ spectrum acquired from the area in \emph{A},
showing kinks at $\sim\pm35\,{\rm mV}$ ($\varepsilon_{{\rm K}}$),
the minimum $\varepsilon_{{\rm min}}$ offset from $\varepsilon_{{\rm F}}$
($V_{{\rm bias}}=0$) by $\sim17\,{\rm mV}$, and marked asymmetry
about $\varepsilon_{{\rm F}}$. Setpoint parameters: sample bias,
$V_{{\rm sample}}=-50\,{\rm mV}$ (\emph{A}), $-60\,{\rm mV}$ (\emph{B}),
$-80\,{\rm mV}$ (\emph{C}); junction resistance, $R_{{\rm J}}=2.5\:{\rm G}\Omega$
(\emph{A}), $0.1\:{\rm G}\Omega$ (\emph{B}), $0.4\:{\rm G}\Omega$
(\emph{C}); and RMS lock-in excitation, $V_{{\rm mod}}=3.5\,{\rm mV}$
(\emph{C}).\label{fig:3qcdw}}
\end{figure*}

\begin{figure*}
\includegraphics[width=6in]{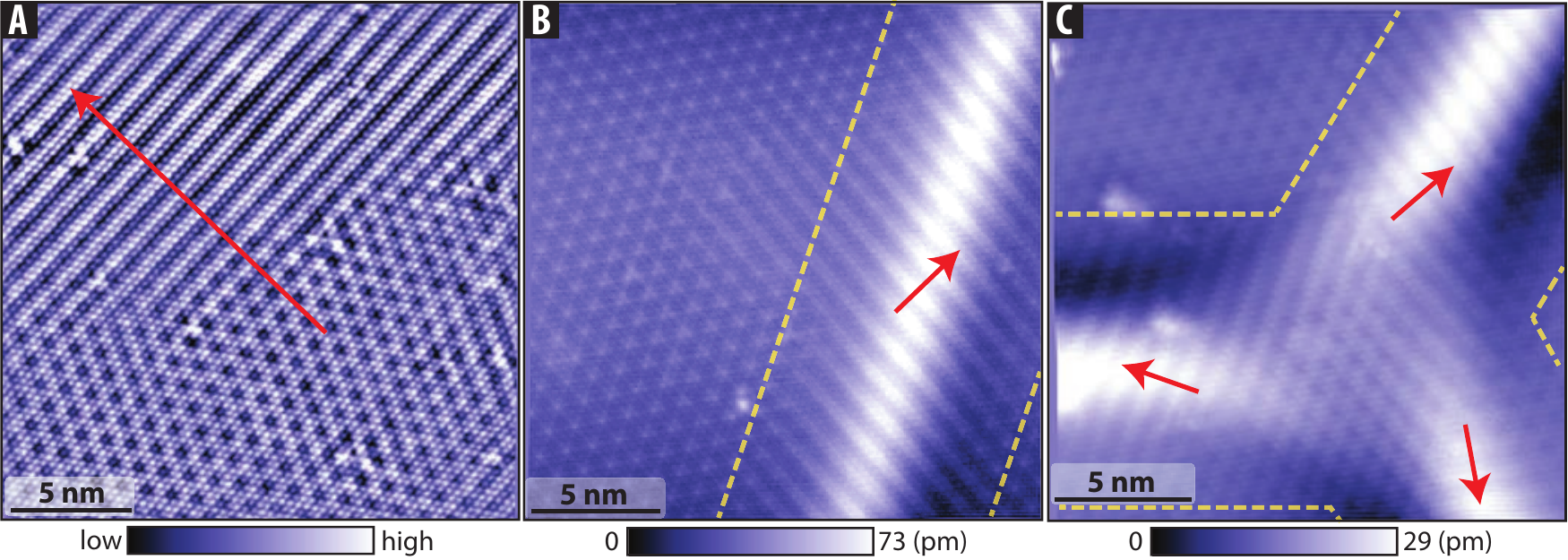}

\caption{\textbf{The $1Q-3Q$ Interface.} (\textbf{\emph{A}}) Topograph showing
an atomically smooth interface between the $3Q$ (bottom) and $1Q$
(top) CDWs. (\textbf{\emph{B}}) Topograph showing formation of $1Q$
CDW along a 'ribbon'. Ribbons are typically $10-20\,{\rm nm}$ in
width, and elevated by $20-40\,{\rm pm}$. Dashed yellow lines indicate
the approximate extent of the ribbon; the red arrow points along the
$1Q$ CDW direction ($\vec{q}_{1Q}$). (\textbf{\emph{C}}) Topograph
showing a $Y$-junction between three such $1Q$ ribbons, each with
a different $1Q$ CDW direction. The topograph in \emph{A} has been
leveled by removing a polynomial background to clearly show the CDW
interface. Setpoint parameters: $V_{{\rm sample}}=-50\,{\rm mV}$
; $R_{{\rm J}}=1\:{\rm G}\Omega$ (\emph{A,C}), $5\:{\rm G}\Omega$
(\emph{B}).\label{fig:1q3qinterface}}
\end{figure*}

\ref{fig:3qcdw}\emph{A }shows a topographic image of a locally commensurate
$(3a_{0}$) CDW. Its microscopic $3Q$ nature is confirmed by the
existence of a secondary CDW peak in the Fourier Transform (FT) in
\ref{fig:3qcdw}\emph{B}, in contrast to bulk measurements\cite{Feng2011}.
Phase slips result in an overall periodicity of $\lambda_{3Q}\simeq3.05\, a_{0}$,
corresponding to $\vec{q}_{3Q}\simeq0.328\,\vec{Q}_{0}$, where $\vec{Q}_{0}$
is the Bragg vector\cite{Moncton1975,Feng2011,McMillan1976}. Our
primary experimental discovery is shown in \ref{fig:1q3qinterface}\emph{A},
where regions of unidirectional ($1Q$) CDW with unique wavevector
$\vec{q}_{1Q}$ along a single $3Q$ direction form an atomically
smooth interface with the $3Q$ CDW. The absence of atomic lattice
discontinuities rules out the possibility of a NbSe$_{2}$ polytype
interface\cite{Wang2009g}. While other TMDCs are known to exhibit
a thermally induced triclinic CDW state that varies with doping near
$T_{{\rm CDW}}$\cite{McMillan1976,Bando1997}, no such anisotropy
has been reported in $2H$-NbSe$_{2}$. Moreover, our observations
are at temperatures $T\ll T_{{\rm CDW}}$, where thermal fluctuations
are insignificant, implying that the $1Q$ CDW is a distinct quantum
phase.

The $1Q$ CDW regions take the form of elevated topographic ribbons,
exemplified in \ref{fig:1q3qinterface}\emph{B-C}, suggesting a strain-induced
phase (\ref{sec:som_ribbons}). The observation of $Y$-junctions
between ribbons with differently oriented $\vec{q}_{1Q}$ rules out
extrinsic uniaxial strain and suggests instead locally varying strain,
perhaps due to underlying defects causing nanoscale buckling of the
top few atomic layers. From a survey of several ribbons, we place
upper bounds of $3\%$ on the vertical strain and $0.06\%$ on the
lateral strain (see \ref{sec:som_ribbons}).

\begin{figure*}
\includegraphics[width=4in]{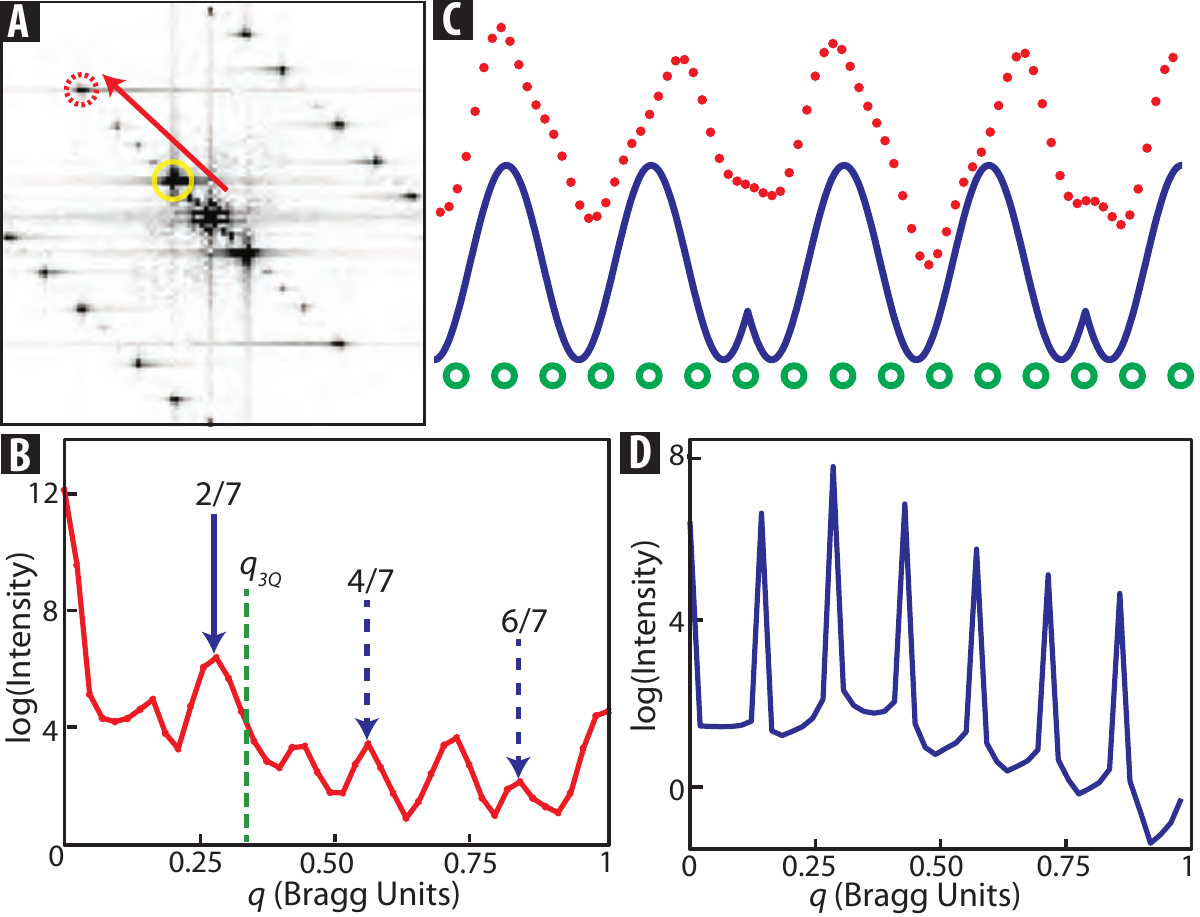}

\caption{\textbf{The $1Q$ CDW}. (\textbf{\emph{A}}) FT of the out-of-feedback
current at $+50\,{\rm mV}$ over a $1Q$ region (upper left quadrant
of \ref{fig:1q3qinterface}\emph{A}). The dominant CDW wavevector
(yellow circle) and Bragg vector (dashed red circle) are indicated.
(\textbf{\emph{B}}) Linecut of the FT intensity parallel to the red
line in \emph{A} from the center to the Bragg peak, in units of the
Bragg vector $\vec{Q}_{0}$. The dominant peak, $q_{1Q}\simeq2/7Q_{0}$
(solid blue arrow) and its harmonics (dashed blue arrows) are identified,
and are distinct from the $3Q$ wavevector $q_{3Q}$ (dashed green
line). The remaining peaks are Bragg reflections of these three peaks.
Setpoint parameters: $V_{{\rm sample}}=-50\,{\rm mV}$; $R_{{\rm J}}=0.2\:{\rm G}\Omega$.
(\textbf{\emph{C}}) Phenomenological model of the observed $1Q$ CDW
wavelength, $\lambda_{1Q}\simeq3.5\, a_{0}$ (details in text). The
atomic periodicity is indicated by green circles and the phase of
the CDW order parameter from the model is shown in blue\cite{McMillan1976}.
A topographic linecut (red dots) is extracted from \ref{fig:1q3qinterface}\emph{A}
along the red arrow ($\vec{q}_{1Q}$), filtered to remove atomic corrugations,
and overlaid for comparison. (\textbf{\emph{D}}) Simulated FT intensity
from the cartoon CDW modulation in \emph{C}, for comparison with experimental
peak positions in \emph{B}.\label{fig:1qcdw}}
\end{figure*}

\ref{fig:1qcdw}\emph{A-B }show the dominant Fourier peak for the
$1Q$ modulation, $\vec{q}_{1Q}\simeq2/7\,\vec{Q}_{0}$, corresponding
to a wavelength, $\lambda_{1Q}\simeq3.5\, a_{0}$. No similar periodicity
or rich harmonic structure has thus far been reported in any TMDC
system\cite{Wilson2001}. We develop a phenomenological understanding
of the $1Q$ harmonic structure following McMillan's Landau theory\cite{McMillan1975,McMillan1976}.
Rather than a uniform $3.5\, a_{0}$ charge modulation, the system
could lower its energy by locking the charge modulation to the lattice
with $3\, a_{0}$ periodicity. This would require compensation by
a one atom phase slip every two oscillations, corresponding to a $2\pi/3$
discommensuration, as shown in \ref{fig:1qcdw}\emph{C}\cite{McMillan1976}.
The resulting harmonic structure shown in \ref{fig:1qcdw}\emph{D}
reproduces all observed peak positions. Moreover, the rich harmonic
content we observe is another indication of the strong coupling of
the electronic modulation to the lattice. An even better agreement
with relative peak heights could be obtained by considering spatial
variations in the order parameter amplitude\cite{McMillan1976}.

\begin{figure}
\includegraphics[width=2.75in]{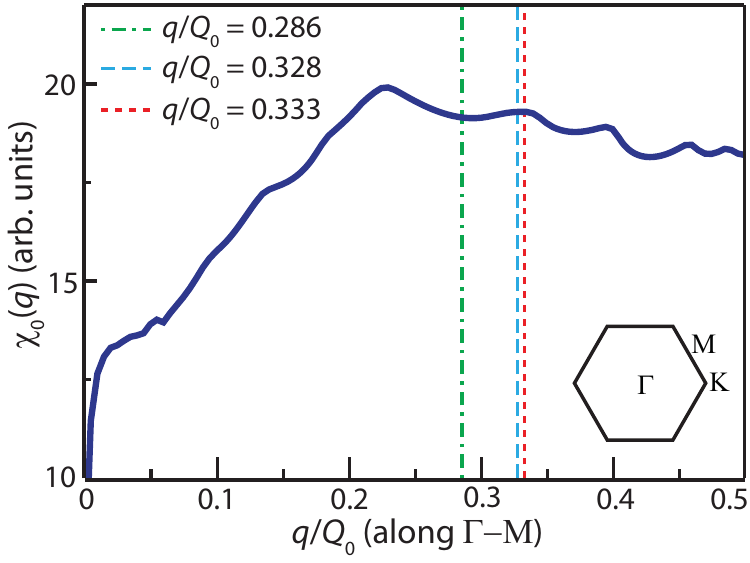}

\caption{\textbf{Susceptibility.} The non-interacting susceptibility $\chi_{0}(q,\omega=0)$
calculated from a fit to the $2H$-NbSe$_{2}$ band structure (\ref{sec:som_bands})
along the $\Gamma-M$ direction of the reciprocal lattice, displaying
a broad maximum over a range of wavevectors: $\tilde{q}\sim0.25-0.4$\cite{Johannes2006,Weber2011,Rahn2012}.
The CDW wavevectors $q_{1Q},\, q_{3Q}$ and $Q_{0}/3$ are overlaid
for comparison. Inset shows the Brillouin Zone (BZ) of the hexagonal
$2H$-NbSe$_{2}$ lattice.\label{fig:susceptibility}}

\end{figure}

The stark contrast between our observation of two CDW wavevectors
$\vec{q}_{1Q}$ and $\vec{q}_{3Q}$ of same orientation but $13\%$
difference in magnitude, and the recent X-ray measurements reported
by Feng \emph{et al.}\cite{Feng2011}, provides strong evidence against
FS nesting at one particular wavevector as a driving force for either
CDW. While our $13\%$ wavevector difference arose from moderate anisotropic
strain (up to $0.06\%$ in-plane), Feng \emph{et al.} applied hydrostatic
pressure sufficient to induce in-plane lattice distortions up to $1.6\%$,
yet observed no measurable deviation of the CDW wavevector from $\vec{q}_{3Q}$\cite{Feng2011}.
The observed insensitivity of $q_{3Q}$ to hydrostatic pressure would
clearly indicate that the FS does not qualitatively change in the
presence of even relatively large lattice distortions, and would thus
rule out a change in the FS as the source of our observed $13\%$
wavevector difference. Furthermore, consistent with our experiment
and with previous calculations\cite{Johannes2006,Weber2011,Rahn2012},
we find no sharp peak in the susceptibility (\ref{fig:susceptibility})
computed from our modeled band structure (\ref{sec:som_bands}). Therefore,
our observations and calculations both indicate that the FS can play
only a minor role in determining CDW wavevectors in NbSe$_{2}$. This
highlights the key role that the $q$-dependence of alternative mechanisms
such as electron-phonon coupling may play in driving the transition,
and particularly the manner in which these mechanisms may be influenced
by local strain.

\begin{figure*}
\includegraphics[width=5.2in]{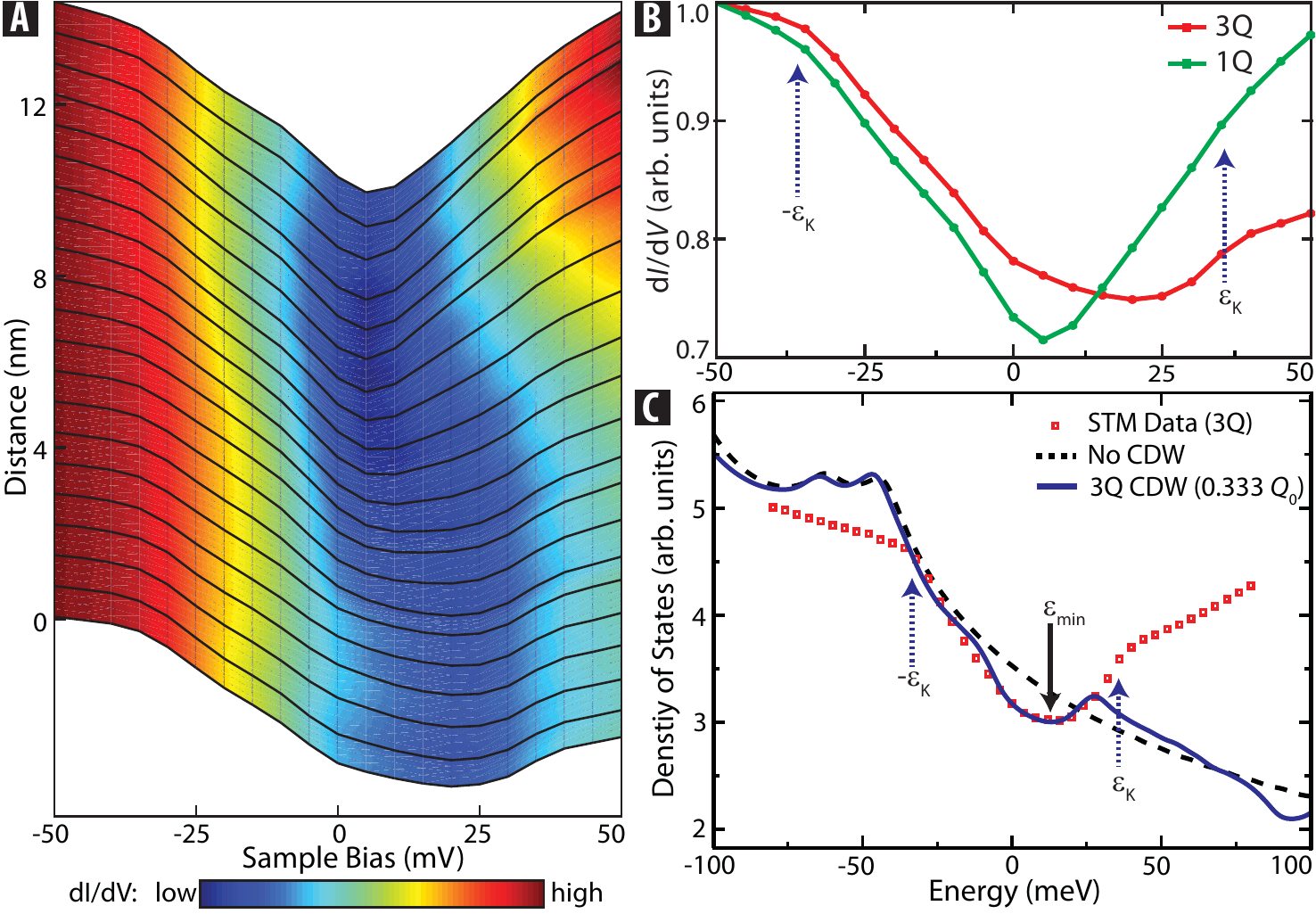}

\caption{\textbf{Spectroscopy across the $1Q-3Q$ Interface.} (\textbf{\emph{A}})
Linecut of $dI/dV$ spectra across the $1Q-3Q$ interface taken along
the red arrow in \ref{fig:1q3qinterface}\emph{A}. (\textbf{\emph{B}})
Representative spectra in the $3Q$ (red) and $1Q$ (green) regions
of \ref{fig:1q3qinterface}\emph{A} overlaid for comparison. The $1Q$
spectra have a minimum close to $\varepsilon_{{\rm F}}$, a deep $V$-shape
with reduced asymmetry, and kinks at $\pm\varepsilon_{{\rm K}}$.
The spectra are normalized at $-50\,{\rm mV}$. Setpoint parameters:
$V_{{\rm sample}}=-50\,{\rm mV}$; $R_{{\rm J}}=0.2\:{\rm G}\Omega$;
$V_{{\rm mod}}=3.5\,{\rm mV}$. (\textbf{\emph{C}}) Calculated DOS
for NbSe$_{2}$ in the `normal' state (black) and in the presence
of a $3Q$ CDW (blue) using $\tilde{q}=0.333\, Q_{0}$, $\tilde{\Delta}=12\,{\rm mV}$,
$\tilde{\Gamma}=5\,{\rm mV}$, compared with the $3Q$ STM spectrum
(red) (details in \ref{sec:som_doscalc}). The calculations reproduce
the observed asymmetry, offset $\varepsilon_{{\rm min}}$, and shape
of the gap structure.\label{fig:spectroscopy}}
\end{figure*}

The two CDW regions display quite different tunneling spectra, as
shown in \ref{fig:spectroscopy}\emph{A-B}. We employ a fit to the
NbSe$_{2}$ band structure (\ref{sec:som_bands}), and impose a CDW
wavevector $\tilde{q}$, gap $\tilde{\Delta}$, and broadening parameter
$\tilde{\Gamma}$, to calculate the density of states (DOS), in \ref{fig:spectroscopy}\emph{C
}(\ref{sec:som_doscalc}). For the $3Q$ CDW state the measured $dI/dV$
spectrum, proportional to the DOS, is best reproduced using $\tilde{q}=(0.333\pm0.004)\: Q_{0}$
(the observed local CDW periodicity), $\tilde{\Delta}=(12\pm2)\,{\rm mV}$
(which has not been previously apparent from direct observations by
spectroscopic techniques), and $\tilde{\Gamma}=5\,{\rm mV}$. The
calculations capture the overall shape, width, and center $\varepsilon_{{\rm min}}$
of the gap structure within 30 mV of $\varepsilon_{{\rm F}}$ (\ref{sec:som_doscalc}).
The fact that $\varepsilon_{{\rm min}}$ is offset from $\varepsilon_{{\rm F}}$
should be unsurprising for a quasi-2D system\cite{Norman2007}, but
had not been understood or observed in NbSe$_{2}$ until now, due
to limitations of spectroscopic techniques which are sensitive to
filled states only\cite{Borisenko2009}.

We disentangle the CDW gap from other effects in the $3Q$ spectra
through a comparison with the $1Q$ spectra in \ref{fig:spectroscopy}\emph{B}.
These $V$-shaped $1Q$ spectra resemble the linear tunneling conductance
background historically attributed to the inelastic coupling of tunneling
electrons to a flat bosonic spectrum\cite{Kirtley1992}. That this
background is much stronger in the $1Q$ region, obscuring band structure
effects, is likely a strain-induced phenomenon, which may be related
to the buckling and associated decoupling of the topmost layers in
the $1Q$ region. Meanwhile, present in both $1Q$ and $3Q$ spectra
(thus unlikely to be associated with these different CDWs), yet absent
in calculations (thus unlikely to be a band structure effect), are
the $\pm35\,{\rm mV}$ kinks, previously and mistakenly identified
as the CDW gap\cite{Hess1991}. We universally observe the kinks even
well above $T_{{\rm CDW}}$, which further demonstrates their lack
of bearing on the CDW phase (\ref{sec:som_doscalc}). ARPES studies
observe a prominent band structure kink at a similar energy in the
Se $\Gamma$-pocket\cite{Valla2004,Rahn2012}, attributed to coupling
to an optical phonon\cite{Mialitsin2010}. We therefore conclude that
this self-energy effect is responsible for the $\varepsilon_{{\rm K}}$
kinks in the tunneling spectra as well. The discrepancy between the
data and band structure calculations above $\sim30\,{\rm mV}$ in
\ref{fig:spectroscopy}\emph{C} can thus be attributed to the inelastic
tunneling background and self-energy effects.

\section{Discussions\label{sec:disc}}

We therefore resolve a longstanding debate about the anomalous CDW
gap magnitude reported by STM measurements\cite{Hess1991}, and caution
that not all $\varepsilon_{{\rm F}}$-symmetric kinks in tunneling
spectra are associated with order (e.g. density wave or superconducting
gaps). On the contrary, we emphasize that the true CDW signature in
NbSe$_{2}$ is offset from $\varepsilon_{{\rm F}}$, which has confused
an active research community for two decades, and has been disentangled
now only by a combination of spatially resolved filled and empty state
spectroscopy of a proximate ($1Q$) phase, and band structure calculations\cite{Norman2007}.
This emphasizes the need for full experimental exploration of proximate
phases in other pertinent materials, combined with quantitative modeling.
We further suggest that controlled local strain, through epitaxy or
intentional defects, may be a useful tuning parameter to access the
necessary proximate phases for comparison.

Beyond providing new insight into the nature of the $3Q$ CDW in NbSe$_{2}$,
our work motivates the utility of the $1Q-3Q$ interface in NbSe$_{2}$
as a platform to explore competing quantum phases in the weakly correlated
limit, as a step towards understanding them in strongly correlated
systems. In the Landau picture of CDWs\cite{McMillan1975}, a quantum
phase transition between $3Q$ and $1Q$ states can arise by tuning
the coefficient of the interaction term between the three inequivalent
CDW propagation directions (though in our case, the magnitude of $q$
differs between the two states). In NbSe$_{2}$, even at low temperatures
$T\ll T_{{\rm CDW}}$, where the amplitude of the order parameter
is already large, moderate strain is seen to have a strong influence,
indicating that the system is intrinsically close to the QCP separating
these states. We note that a related phase transition between the
observed $1Q$ CDW phase and a `hidden' $2Q$ phase has been suggested,
but not directly visualized, in the rare-earth tritellurides\cite{Yao2006,Ru2008a}.

Our discovery provides a new perspective on the role of density wave
order in complex systems. First, our resolution of two longstanding
debates about NbSe$_{2}$ puts this much-studied material on firmer
footing as a well-understood model system for CDW studies and competing
ground states in superconductors. We have disentangled the true CDW
gap, and clarified that FS nesting plays a minor role in determining
the CDW wavevectors in this material, thereby highlighting the role
of other mechanisms in driving the transition. Second, our revelation
of a particle-hole asymmetric CDW gap emphasizes the limitations of
filled-state-only probes, e.g. ARPES, for investigating phases other
than SC - which is unambiguously particle-hole symmetric. Full spectral
probes such as STM, in combination with quantitative calculations,
are necessary to understand the competition between SC and particle-hole
asymmetric phases. Third, our observation of the local effect of even
moderate strain in driving a quantum phase transition calls for a
reinvestigation of possible phase inhomogeneity in other strongly
correlated systems, where larger strain may occur\cite{Slezak2008,Chu2010a}.

In the cuprate superconductors, an analogous phase boundary between
unidirectional ($1Q$) charge `stripes' and bidirectional ($2Q$)
`checkerboard' has been predicted\cite{Robertson2006,DelMaestro2006a}.
The introduction of quenched disorder results in discommensurations
in the $2Q$ phase and disordered orientational order in the $1Q$
phase, making them hard to distinguish - especially in the cuprate
BSCCO, thought to be in proximity to the $1Q-2Q$ phase boundary\cite{Robertson2006,DelMaestro2006a}.
Recent STM studies of the $\sim4\, a_{0}$ charge order in BSCCO have
had conflicting interpretations, with independent suggestions of fluctuating
$2Q$ and $1Q$ order\cite{Wise2008,Parker2010}. However, the influence
of strain, from the supermodulation lattice buckling, or from randomly
distributed dopants, is seldom accounted for. Previous studies have
shown that both these strain phenomena correlate with nanoscale electronic
inhomogeneity\cite{Slezak2008,Zeljkovic2012b}. A possible explanation
is local stabilization of the $1Q$ state, producing $1Q-2Q$ and
$1Q-1Q$ interfaces with spectral differences, analogous to \ref{fig:1q3qinterface}
and \ref{fig:spectroscopy}\emph{A-B}. While the presence of strong
disorder (up to $12\%$ strain variations on a nanometer length scale\cite{Slezak2008})
complicates the interpretations in BSCCO, we stress the importance
of isolating and modeling strain effects for better understanding
and control of the phase transitions in cuprates. Finally, the microscopic
visualization of the role of strain in stabilizing new order suggests
a controlled route towards engineering novel quantum phases and interfaces
and studying symmetry breaking in strongly correlated materials. In
this regard, we suggest a connection to the emerging importance of
strain as a route to high-$T_{{\rm C}}$ superconductivity in novel
iron-based materials\cite{Saha2012,Wang2012b}.

\section{Methods\label{sec:methods}}

\paragraph*{\textbf{STM Experiments.}}

Measurements were performed using a home-built STM at temperatures
between $2-10$ K. Magnetic fields of up to $5\:{\rm T}$ were used
to suppress the superconducting state as needed. Single crystals of
$2H$-NbSe$_{2}$ were cleaved in situ in cryogenic ultrahigh-vacuum
and inserted into the STM. A mechanically cut PtIr tip, cleaned by
field emission and characterized on gold, was used for the measurements.
Spectroscopy data were acquired using a lock-in technique at 1.115~kHz.
The topographic and spectroscopic signatures of the $1Q$ ribbons
have been verified with several tips.

\paragraph*{\textbf{Band Structure and DOS Calculations.}}

The band structure of $2H$-NbSe$_{2}$ close to $\varepsilon_{{\rm F}}$
consists of two Nb$-4d$ derived bands and one Se$-4p$ `pancake'-shaped
hole pocket\cite{Johannes2006}. The Nb$-4d$ bands are modeled using
a tight-binding fit to the observed ARPES band structure\cite{Rahn2012},
and the Se$-4p$ pocket is approximated by a simple quadratic form
to fit LDA calculations\cite{Johannes2006} (details in \ref{sec:som_bands}).
Using all three bands, the DOS in the presence of a CDW is calculated
by imposing a constant coupling between electronic states connected
by any one of the three $q$-vectors. The strength of the coupling
$\tilde{\Delta}$ is taken as a free parameter in the reproduction
of the experimentally observed DOS, and the size of $\tilde{q}$ is
allowed to vary slightly around the observed value of $0.328\, Q_{0}$
(details in \ref{sec:som_doscalc}).

\rule[0.5ex]{0.8\columnwidth}{1pt}

\paragraph*{\textbf{Acknowledgements.}}

We are grateful to Patrick Lee and Steve Kivelson for insightful discussions
and to Wilfried Krüger, Lutz Kipp and Daniel Walkup for useful experimental
inputs. This work was supported by NSF DMR-0847433 and the New York
Community Trust - George Merck Fund (Harvard), DOE, Office of Science,
under Contract No. DE-AC02-06CH11357 (Argonne) and DFG via SFB 855
(Kiel). In addition, we acknowledge funding from A{*}STAR, Singapore
(A.S.) and NSERC, Canada (M.M.Y.).

\paragraph*{\textbf{Author Contributions.}}

A.S., M.M.Y. and Y.H. performed STM experiments, and A.S. led the
data analysis with contributions from M.M.Y. D.J.R. and K.R. grew
and characterized samples. J.v.W., K.R. and M.R.N. performed calculations.
A.S. wrote the main manuscript with theoretical sections written by
J.v.W. E.W.H., K.R., M.R.N. and J.E.H. advised the work and edited
the manuscript.

\paragraph*{\textbf{Author Information.}}

The authors declare no competing financial interests. Correspondence
and requests for materials should be addressed to A.S. (anjan@physics.harvard.edu)
and J.E.H. (jhoffman@physics.harvard.edu).

\newpage
\linespread{1.1}
\setlength{\parskip}{2ex}

\renewcommand{\thesection}{SI~\Roman{section}}
\renewcommand{\theequation}{S\arabic{equation}}
\renewcommand{\thetable}{S\arabic{table}}
\setcounter{section}{0}

\renewcommand{\thefigure}{S\arabic{figure}}
\renewcommand\theHfigure{S\arabic{figure}}
\setcounter{figure}{0}

\onecolumngrid
\begin{center}
\textsc{\textbf{\Large Supporting Information}}
\end{center}
\vspace{6ex}
\twocolumngrid

\section{Sample Characterization\label{sec:som_samplechar}}

\begin{figure}[h]
\includegraphics[width=3in]{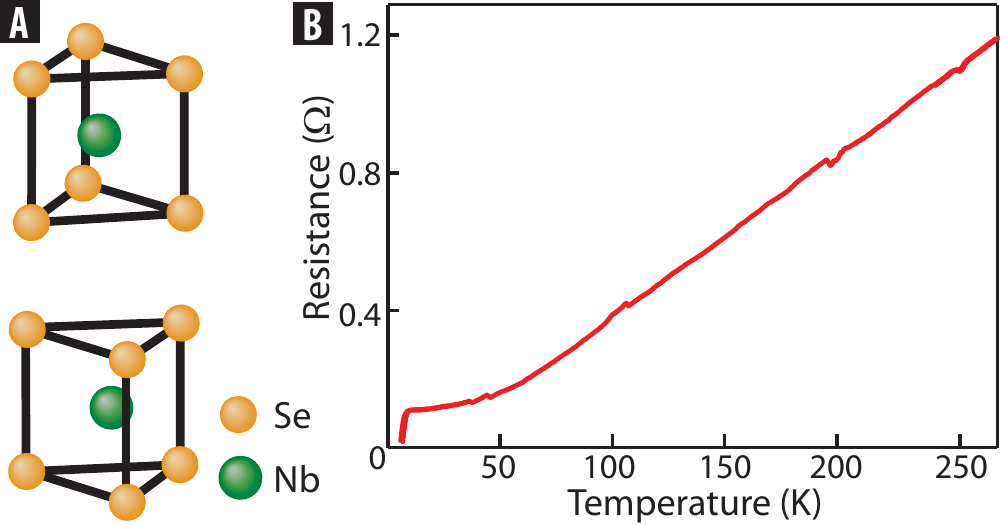}

\caption{\textbf{Sample Characterization.} (\textbf{\emph{A}}) The layered
hexagonal crystal structure of $2H$-NbSe$_{2}$, with alternating
sandwiches of Se-Nb-Se. (\textbf{\emph{B}}) Temperature dependence
of the resistance for the sample batch used for this study, showing
a superconducting transition at $\sim7\,{\rm K}$.\label{fig:som_samplechar}}
\end{figure}

$2H$-NbSe$_{2}$ is a layered transition metal dichalcogenide with
a hexagonal structure and $D_{6h}^{4}$ space group symmetry. The
unit cell (\ref{fig:som_samplechar}\emph{A}) consists of two sandwiches
of Se-Nb-Se. The crystal typically cleaves between the neighboring
Se layers, coupled by weak van der Waals forces.

Single crystals of $2H$-NbSe$_{2}$ were grown by chemical vapor
transport using iodine as the transport agent. A transport characterization
of the sample batch used in this work is shown in \ref{fig:som_samplechar}\emph{B}.
The superconducting transition is observed at $T_{{\rm SC}}\sim7{\rm \, K}$.
The residual resistivity ratio (RRR), defined as the ratio of resistances
$R(295\,{\rm K})/R(7.5\,{\rm K})$, is $\sim16$.

\section{$1Q$ 'Ribbons': Height and Orientation\label{sec:som_ribbons}}

\begin{figure*}
\includegraphics[width=5.4in]{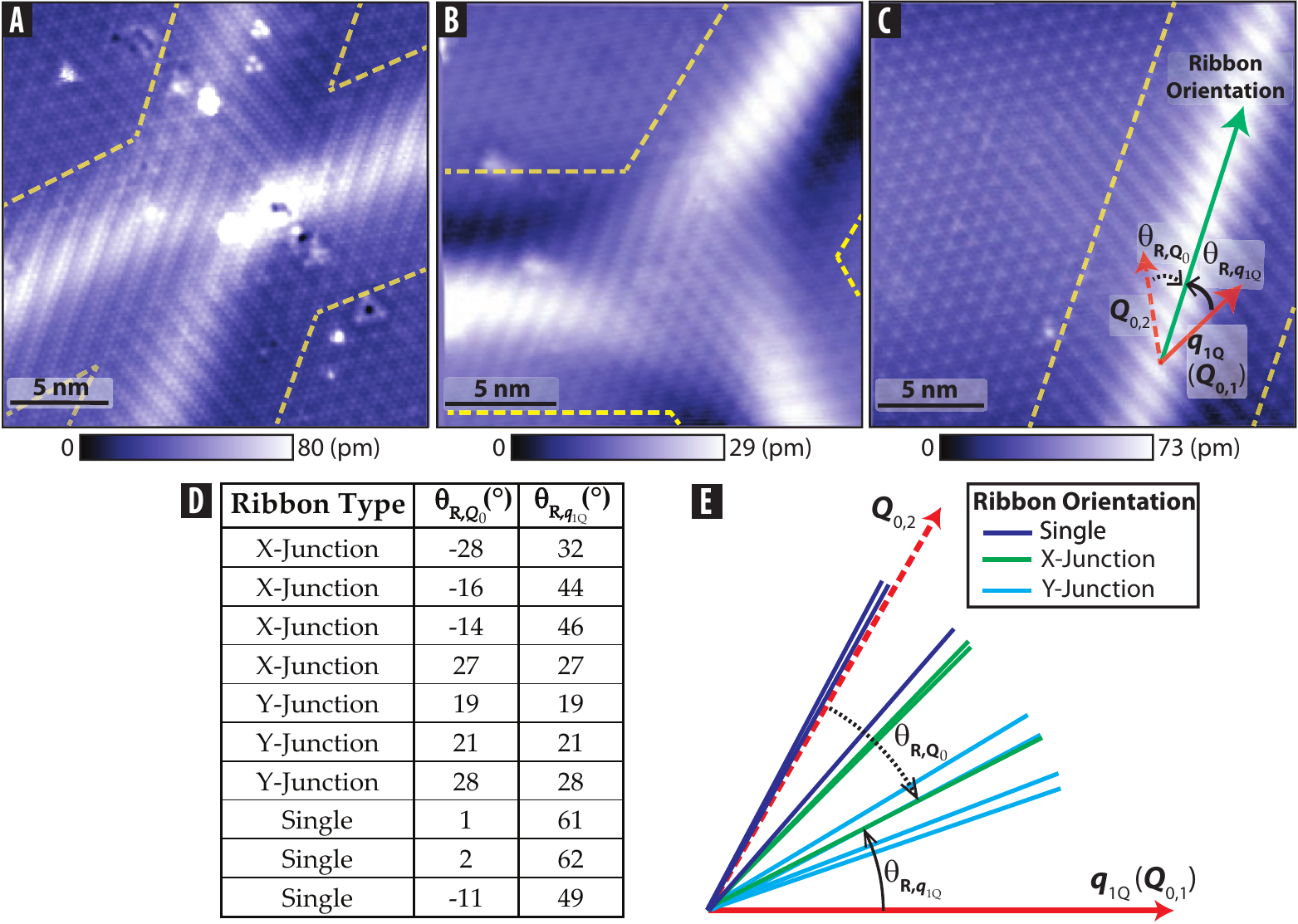}

\caption{\textbf{1Q 'Ribbons': Orientation.} (\textbf{\emph{A}},\textbf{\emph{
B}}) STM topographs showing 'ribbons' of unidirectional ($1Q$) CDW,
intersecting to form an $X$-junction (\emph{A}) and a $Y$-junction
(\emph{B}), with CDW wavevector $\vec{q}_{1Q}$ varying between the
arms. Dashed yellow lines indicate the approximate extent of the ribbons.
Setpoint parameters: $V_{{\rm sample}}=-50\,{\rm mV}$ for both; $R_{{\rm J}}=5\:{\rm G}\Omega$
(\emph{A}), $1\:{\rm G}\Omega$ (\emph{B}). (\textbf{\emph{C}}) STM
topograph of a $1Q-3Q$ interface (top arm of $X$-junction, also
shown in \ref{fig:1q3qinterface}\emph{B}), demonstrating the definitions
of the relative orientations $\theta_{{\rm R},Q_{0}}$, between the
ribbon (green arrow) and the nearest Bragg vector (dashed red arrow),
and $\theta_{{\rm R},q_{1Q}}$, between the ribbon and $1Q$ wavevector
$\vec{q}_{1Q}$ (solid red arrow). (\textbf{\emph{D}}) A table detailing
the values of $\theta_{{\rm R},Q_{0}}$ and $\theta_{{\rm R},q_{1Q}}$
observed in the various $1Q$ ribbons studied in this work, with the
first entry corresponding to \emph{C}. (\textbf{\emph{E}}) A visual
illustration of the spread of values in table \emph{D}. The dark blue
(single), green ($X$-junction) and cyan ($Y$-junction) lines describe
the orientation of the various ribbons with respect to the $1Q$ wavevector
$\vec{q}_{1Q}$ along the $\vec{Q}_{0,1}$ Bragg vector (solid red
arrow) and another Bragg vector $\vec{Q}_{0,2}$ (dashed red arrow).\label{fig:som_ribbonorient}}
\end{figure*}

The $1Q$ CDW typically appears in regions which persist in one direction
with apparent $20-40\,{\rm pm}$ topographic elevation, forming a
$10-20\,{\rm nm}$ wide 'ribbon' structure (\ref{fig:som_ribbonorient}\emph{A-C}).
We note that the topographic elevation $z_{{\rm STM}}(\vec{r},V_{0},I_{0})$
as measured by maintaining a constant current $I_{0}$ with bias setpoint
$V_{0}$ at lateral tip position $\vec{\ensuremath{r}}\equiv(x,y)$
can be given by

\begin{equation}
z_{{\rm STM}}(\vec{r},V_{0},I_{0})\simeq z_{{\rm T}}(\vec{r})+\frac{1}{\kappa(\vec{r})}\cdot\ln\left(\frac{I_{0}}{\int_{0}^{V_{0}}{\rm d}V\: eD\left(\vec{r},eV\right)}\right)
\end{equation}

Here $z_{T}(\vec{r})$ is the true topographic corrugation of the
sample, $\kappa(\vec{r})$ is a measure of the local tunnel barrier
height (LBH), and $D\left(\vec{r},eV\right)$ is the local density
of states (LDOS) of the sample at energy $eV$. Because of the logarithmic
sensitivity of $z_{{\rm STM}}(\vec{r},V_{0},I_{0})$ to the integral
of the LDOS from the Fermi energy, $\varepsilon_{{\rm F}}$ (corresponding
to $V=0$) up to the bias setpoint $eV_{0}$, STM topographs may contain
electronic artifacts masquerading as geometric effects. Therefore,
we present two pieces of evidence for the true geometric elevation
of these ribbons.

First, a tabulation of the relative orientation $\theta_{{\rm R},Q_{0}}$
of the ribbon to the nearest Bragg vector of the underlying hexagonal
lattice for the various ribbons imaged in the study shows a seemingly
random spread from $-30^{\circ}$ to $30^{\circ}$ - the full range
of available angles (\ref{fig:som_ribbonorient}\emph{D-E}). Furthermore,
these ribbon structures can intersect to form $X$ as well as $Y$
junctions (\ref{fig:som_ribbonorient}\emph{A-B}), and the angle between
intersecting ribbons varies from $40^{\circ}$ to $60^{\circ}$. The
fact that ribbon orientation does not respect lattice symmetry strongly
suggests a true geometric, rather than electronic origin of their
apparent height. We contrast this observation with enhanced STM topographic
corrugation associated with predominantly electronic features in a
wide range of other materials, which respect the symmetry of the hexagonal\cite{Hanaguri2010,Rutter2007a}
or square\cite{Wise2008} lattice.

\begin{figure*}
\includegraphics[width=4in]{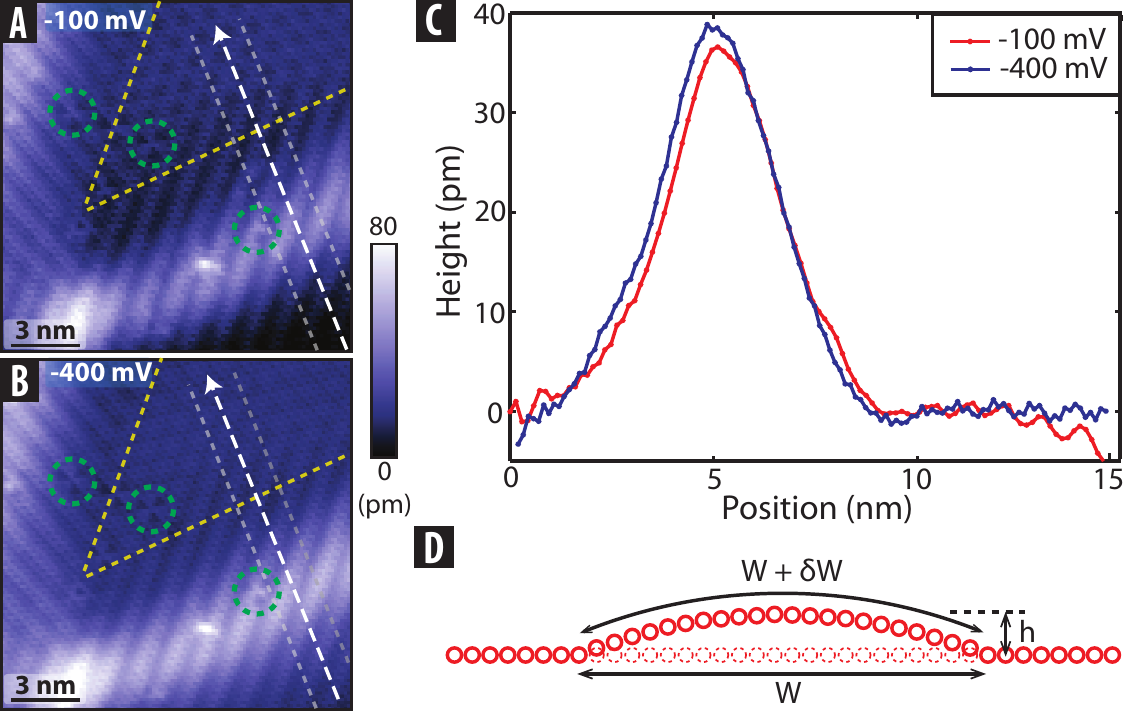}

\caption{\textbf{$1Q$ 'Ribbons': Bias Setpoint Dependence.} (\textbf{\emph{A,
B}}) Topographs of $1Q$ ribbons over the same spatial area acquired
with different bias setpoints: -100\textasciitilde{}mV (\emph{A})
and -400 mV (\emph{B}), with $R_{{\rm J}}=10\:{\rm G}\Omega$ in both.
The dashed yellow lines indicate the approximate extent of the ribbons,
and the dashed green circles enclose triangular impurities, visible
in \emph{B} with a $50-70\%$ larger apparent height than in \emph{A}.
(\textbf{\emph{C}}) Linecuts taken through the topographs in \emph{A}
(red) and \emph{B} (blue) transverse to the lower ribbon, along the
dashed white arrows. The measured ribbon height varies less than $5\%$
between bias setpoints -100 mV and -400 mV. The linecuts have been
laterally averaged over a 3 nm width indicated by the dashed grey
lines. (\textbf{\emph{D}}) A cartoon representation of the lattice
distortion caused by the formation of such a ribbon, modeled as a
half period of a sinusoid. The red circles correspond to rows of displaced
Se atoms, while the dashed red circles represent their original undistorted
positions. The ribbon has height $h$ (40 pm) and width $W$ (10 nm),
resulting in a total lateral distortion of $\delta W\sim45\:{\rm pm}$
across the ribbon.\label{fig:som_ribbonht}}
\end{figure*}

Second, the measured height of these ribbons in STM topographs exhibits
$<5\%$ dependence on bias setpoint within 400 mV below the Fermi
energy (\ref{fig:som_ribbonht}\emph{A-C}). We note that this energy
range over which the measured height of the ribbons is invariant is
much larger than the spectral range of CDW variation in the DOS ($\sim50$
mV). We further note the contrast between the bias-independent ribbons,
and single atom impurity resonances, whose measured 'height' varies
by $50-70\%$ between \ref{fig:som_ribbonht}\emph{A} and \emph{B}.
Therefore we conclude that the measured height ($20-40$ pm) and width
($10-20$ nm) of these ribbons has a predominantly geometric origin.

Having established the topographic origin of these ribbons (\ref{fig:som_ribbonorient}
and \ref{fig:som_ribbonht}), we suggest that these ribbons are likely
a topographic rippling of the top few layers. These ribbons may arise
during the cleaving process due to underlying growth defects which
can intercalate between Se-Nb-Se sandwich layers. We note that similar
topographic ribbon deformations have recently been observed in another
layered chalcogenide (Bi$_{2}$Te$_{3}$)\cite{Okada2012c}.

We estimate the in-plane and out-of-plane lattice strain associated
with the topographic ribbon features. Using the maximum topographic
elevation of an observed ribbon (40 pm), we can put an upper limit
on the out-of-plane distortion by assuming that a minimum of 2 sandwich
layers are elevated (any fewer, and the defects causing the elevation
would be likely visible in our topographs). The out of plane distortion
is therefore $\lesssim3\%$ ($40\,{\rm pm}/12.54\,{\rm \mathring{A}}$)
of the unit cell spacing. To measure the in-plane distortion, we first
use the Lawler-Fujita algorithm\cite{Lawler2010}, which can determine
the lateral location of atoms with precision $\sim2\%$ of the lattice
spacing\cite{Hamidian2012,Williams2011a}. With this algorithm, we
do not observe any change in the lattice constant across the ribbon,
which places a direct experimental upper limit on the in-plane distortion
of $\sim2\%$.

However, we can estimate the actual in-plane distortion indirectly
from the measured out-of-plane distortion, by modeling the ribbon
as a half-period of a sinusoid with height $h$ (40 pm) and width
$W$ (10 nm) (\ref{fig:som_ribbonht}\emph{D}). The total lateral
deformation due to such a ribbon is $\delta W\sim45\,{\rm pm}$, corresponding
to $\sim0.06\%$ of the lattice spacing. As this is well below the
resolution of the Lawler-Fujita algorithm, it is not surprising that
the in-plane distortion is not detectable in STM topographs. From
the upper bounds of $3\%$ on the vertical strain and $0.06\%$ on
the lateral strain, we note that the magnitude of the strain field
leading to the formation of these ribbons is moderate, in comparison
to some other correlated materials\cite{Slezak2008,Chu2010a}. We
also note that while the magnitude of lattice distortion of these
ribbons may seem small in the context of the observed quantum phase
inhomogeneity, a comparison with other known materials suggests that
strain of this magnitude can be sufficient to drive a transition to
the unidirectional CDW phase\cite{Ru2008a}.

We previously discussed the ribbon orientation with respect to the
lattice ($\theta_{{\rm R},Q_{0}}$); we now consider the ribbon orientation
with respect to the $1Q$ CDW wavevector ($\theta_{{\rm R},q_{1Q}}$),
also detailed in \ref{fig:som_ribbonorient}\emph{D}-\emph{E}. In
a simple picture of the strained ribbon structure, we would expect
the ribbon-induced strain to couple strongly to the $1Q$ CDW orientation
either parallel or perpendicular to the ribbon, and thus we would
expect to observe values of $\theta_{{\rm R},q_{1Q}}$ either between
$0-30^{\circ}$ or between $60-90^{\circ}$. Yet we often find $\theta_{{\rm R},q_{1Q}}$
to be in the $30-60^{\circ}$ range as well. Insufficient statistics
prevent us from inferring a clear connection between ribbon orientation
and $\vec{q}_{1Q}$ orientation, but the wide distribution of relative
angles suggests the complexity of the interaction.

\section{Band Structure Calculations\label{sec:som_bands}}

\begin{figure*}
\includegraphics[width=5in]{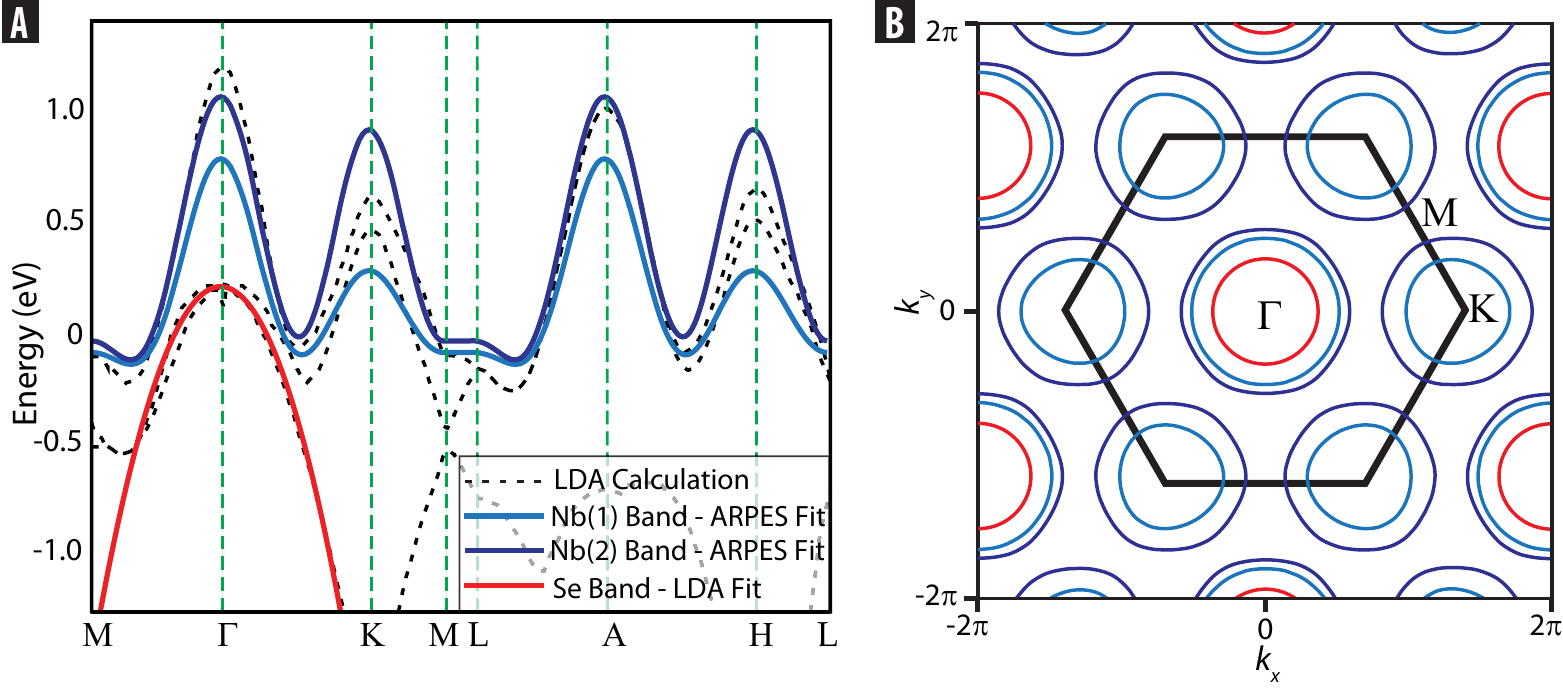}

\caption{\textbf{Band Structure Calculations.} (\textbf{\emph{A}}) The calculated
band structure of $2H$-NbSe$_{2}$ along high-symmetry directions,
showing the two Nb-$4d$ bands obtained using an ARPES tight-binding
fit\cite{Rahn2012}, and the Se-$4p$ band, modeled as a parabolic
fit to LDA calculations of Johannes \emph{et al.}\cite{Johannes2006},
compared with the LDA calculations (dashed black lines). (\textbf{\emph{B}})
The Fermi surface obtained using this band structure fit, with the
BZ shown in black.\label{fig:som_bands}}
\end{figure*}

The band structure of $2H$-NbSe$_{2}$ close to $\varepsilon_{{\rm F}}$
consists of two Nb-$4d$ derived bands and one Se-$4p$ 'pancake'-shaped
hole pocket\cite{Johannes2006,Rossnagel2001}. Close to $\varepsilon_{{\rm F}}$,
the Se-$4p$ pancake-shaped hole pocket which surrounds the $\Gamma$-point
can be modeled by a simple quadratic form,

\begin{equation}
E=A\cdot\frac{a_{0}^{2}}{4\pi^{2}}\,(k_{x}^{2}+k_{y}^{2})+B
\end{equation}

With the values of $A=-5.4\,{\rm eV}$ and $B=-0.65\,{\rm eV}$, this
model accurately reproduces the dispersion obtained in LDA calculations
by Johannes \emph{et al.}\cite{Johannes2006}, as shown in \ref{fig:som_bands}\emph{A}.
To model the Nb-$4d$ bands, we use a tight-binding fit to the band
structure observed in ARPES experiments, as reported by Rahn \emph{et
al.}\cite{Rahn2012}. We find that a small ($\sim+16\,{\rm meV}$)
offset in the chemical potential relative to the parameters used by
Rahn \emph{et al.} was needed to reproduce the observed DOS from STM
measurements. This offset is within the accuracy of the tight-binding
fitting scheme\cite{Rahn2012}. The band energies of the tight-binding
description are given by\cite{Rahn2012}

\begin{align}
E=t_{0} & +t_{1}[2\cos\xi\cos\eta+\cos2\xi]\nonumber \\
 & +t_{2}[2\cos3\xi\cos\eta+\cos2\eta]\nonumber \\
 & +t_{3}[2\cos2\xi\cos2\eta+\cos4\xi]\\
 & +t_{4}[\cos\xi\cos3\eta+\cos5\xi\cos\eta+\cos4\xi\cos2\eta]\nonumber \\
 & +t_{5}[2\cos3\xi\cos3\eta+\cos6\xi]\nonumber
\end{align}

where $\xi=\frac{1}{2}k_{x}a_{0}$ and $\eta=\frac{1}{2}\sqrt{3}k_{y}a_{0}$
and $k_{y}$ is along $\Gamma-M$. The values of the tight binding
parameters (including the offset) used in this work are:

\begin{table}[H]
\begin{centering}
\begin{tabular}{|c|c|c|}
\hline
\textbf{Parameter} & \textbf{Nb Band (1)} & \textbf{Nb Band (2)}\tabularnewline
\hline
\hline
$t_{0}$ & 26.9 & 219.0\tabularnewline
\hline
$t_{1}$ & 86.8 & 46.0\tabularnewline
\hline
$t_{2}$ & 139.9 & 257.5\tabularnewline
\hline
$t_{3}$ & 29.6 & 4.4\tabularnewline
\hline
$t_{4}$ & 3.5 & -15.0\tabularnewline
\hline
$t_{5}$ & 3.3 & 6.0\tabularnewline
\hline
\end{tabular}
\par\end{centering}

\caption{Tight-binding parameters for the two Nb bands (in meV).}
\label{tab:band_parameters}
\end{table}

\section{Density of States Calculations\label{sec:som_doscalc}}

\begin{figure*}
\includegraphics[width=5.1in]{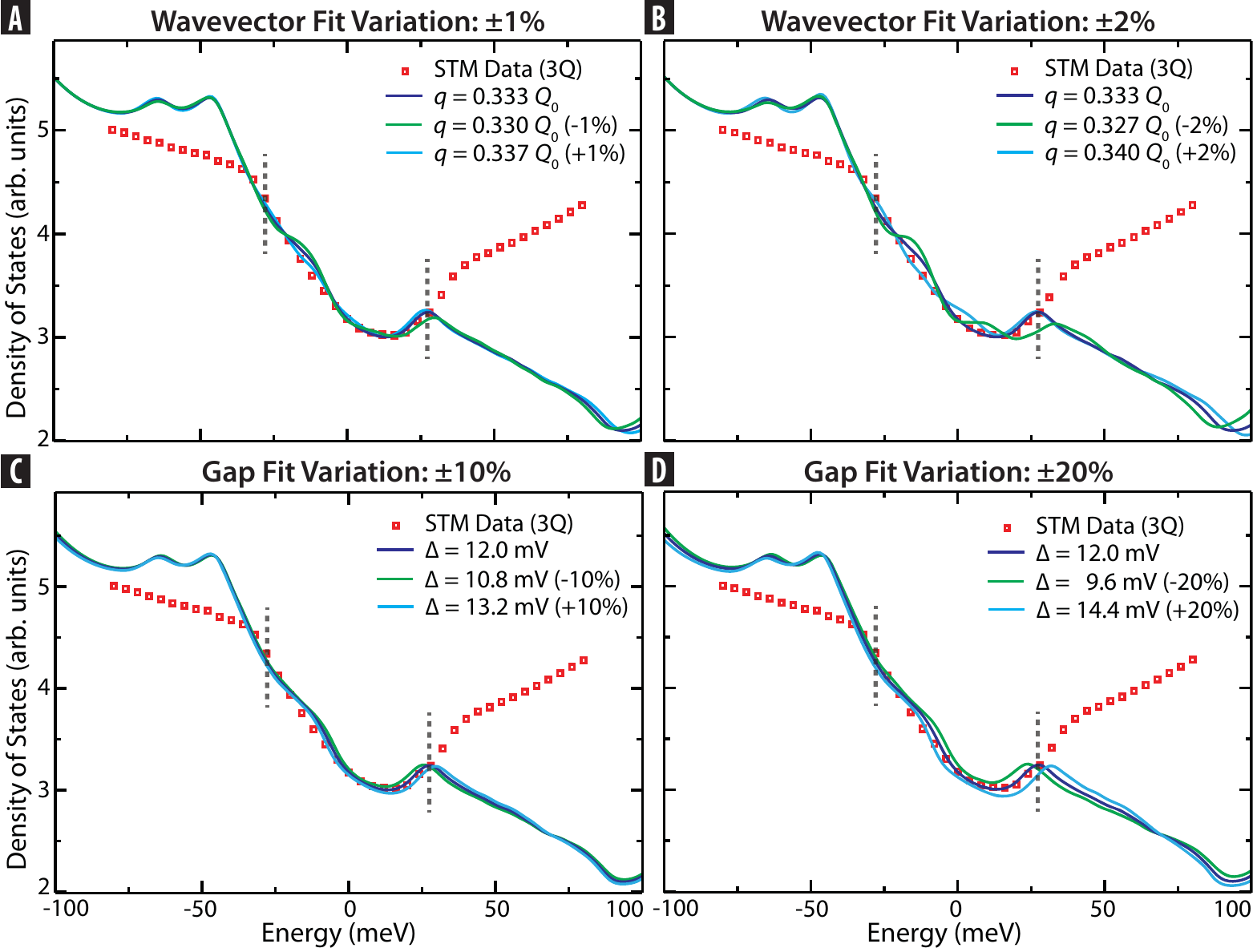}

\caption{\textbf{Variations in DOS Fit Parameters.} Calculated DOS spectrum
using the band structure fit in the presence of a $3Q$ CDW, showing
the effects of varying the fit parameters, wavevector $\tilde{q}$
and gap value $\tilde{\Delta}$ around the best fit values ($\tilde{q}=0.333\, Q_{0}$,
$\tilde{\Delta}=12\,{\rm meV}$, dark blue), compared with the STM
data (red). (\textbf{\emph{A}},\textbf{\emph{ B}}) show the effects
of varying the wavevector $\tilde{q}$ around $0.333\, Q_{0}$ by
$\pm1\%$ and $\pm2\%$ respectively. (\textbf{\emph{C}}, \textbf{\emph{D}})
show the effects of varying the gap value $\tilde{\Delta}$ around
12 meV by $\pm10\%$ and $\pm20\%$ respectively. Error bars for $\tilde{q}$
($0.004\, Q_{0}$) and $\tilde{\Delta}$ (2 meV) are deduced using
these variations.\label{fig:som_dosfiterror}}
\end{figure*}
To calculate the DOS in the presence of a $3Q$ CDW, we impose a coupling
between states connected by any one of the three $\tilde{\vec{q}}$-vectors,
given by:

\begin{equation}
\mathcal{H}_{{\rm CDW}}=\tilde{\Delta}\cdot\sum\limits _{\vec{k}}\,\left(c_{\vec{k}+\tilde{\vec{q}}}^{\dagger}\, c_{\vec{k}}+{\rm h.c.}\right)
\end{equation}
The strength $\tilde{\Delta}$ of the coupling is taken as a free
parameter in the reproduction of the experimentally observed DOS,
with a broadening parameter fixed at $\Gamma=5\,{\rm meV}$. Adjusting
the size of $\tilde{q}$ slightly around the observed value of $q_{3Q}=0.328\, Q_{0}$,
we find the best match with STM spectra using $\tilde{q}=0.333\, Q_{0}$
(corresponding to the locally commensurate CDW periodicity), $\tilde{\Delta}=12\,{\rm meV}$.
With these parameter values, the gap structure in the calculated DOS
closely approaches the overall shape, width and center of the gap
structure seen in the measured data within $\pm30{\rm \, meV}$ of
the Fermi energy $\varepsilon_{{\rm F}}$, as shown in \emph{\ref{fig:spectroscopy}C}.

To demonstrate the accuracy of this fit, we show the effects of varying
the wavevector $\tilde{q}$ by $1-2\%$ (\ref{fig:som_dosfiterror}\emph{A-B}),
and the gap value $\tilde{\Delta}$ by $10-20\%$ (\ref{fig:som_dosfiterror}\emph{C-D}).
Using these fit parameter variations, we estimate the errors for $\tilde{q}$
and $\tilde{\Delta}$ to be $0.004\, Q_{0}$ and $2\,{\rm meV}$ respectively.
The value of $\tilde{\Delta}$ may however be an overestimate, leading
to a systematic error of the same order as the fit uncertainty, since
the described procedure does not take into account the particle-hole
symmetric inelastic background in the experimental DOS. Accurate modeling
of the inelastic background would require detailed temperature dependent
spectroscopic data, which is beyond the scope of this work.

Crucially, we note that the particle-hole asymmetry in the CDW gap,
with its minimum centered above $\varepsilon_{{\rm F}}$, cannot be
removed by the subtraction of a particle-hole symmetric background.
Likewise, the striking deviation of our fitted gap parameter ($\tilde{\Delta}=12\,{\rm meV}$)
from previous results (four times larger than the 3\ meV value detailed
by Borisenko \emph{et al.}\cite{Borisenko2009}, and three times smaller
than the 35\ meV value detailed by Hess \emph{et al.}\cite{Hess1991})
far exceeds fit or systematic uncertainties.

\begin{figure}[h]
\includegraphics[width=2.5in]{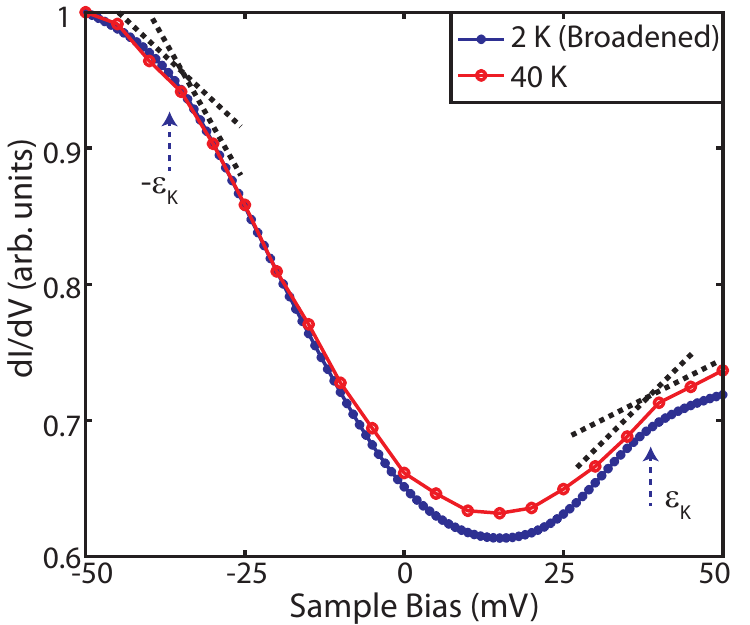}

\caption{\textbf{Spectroscopy above $T_{{\rm CDW}}$}. Spectrum acquired at
$40\,{\rm K}$ (red) compared with that acquired at $2\,{\rm K}$,
$6\,{\rm T}$ (blue) on the same cleaved surface (within $300\,{\rm nm}$).
The $2\,{\rm K}$ spectrum has been thermally broadened to $40\,{\rm K}$
for ease of comparison. The spectral kinks at $\pm35\,{\rm mV}$ ($\pm\varepsilon_{{\rm K}}$)
are distinguishable well above \textbf{$T_{{\rm CDW}}$}, as shown
by the guides to the eye (c.f. \emph{\ref{fig:3qcdw}C} for a low
temperature comparison).\label{fig:som_tdepspectra}}
\end{figure}

To experimentally verify the lack of bearing of the $35\,{\rm mV}$
kinks on the CDW phase, we performed spectroscopy up to $45\,{\rm K}$,
and universally observed the presence of these kinks in the STM spectra
well above \textbf{$T_{{\rm CDW}}$} ($\sim33\,{\rm K}$). A comparison
of the typical spectrum acquired at $40\,{\rm K}$ to that acquired
at $2\,{\rm K}$, is shown in \ref{fig:som_tdepspectra}. We note
that the data acquired at $40\,{\rm K}$ are thermally smeared by
$\mathcal{O}(3k_{B}T)$, i.e. $\sim10\,{\rm mV}$, resulting in a
broadening of the kinks. Despite this, the kinks remain distinguishable,
and are present throughout all spatial regions studied in this work.

\begin{figure*}
\includegraphics[width=5.1in]{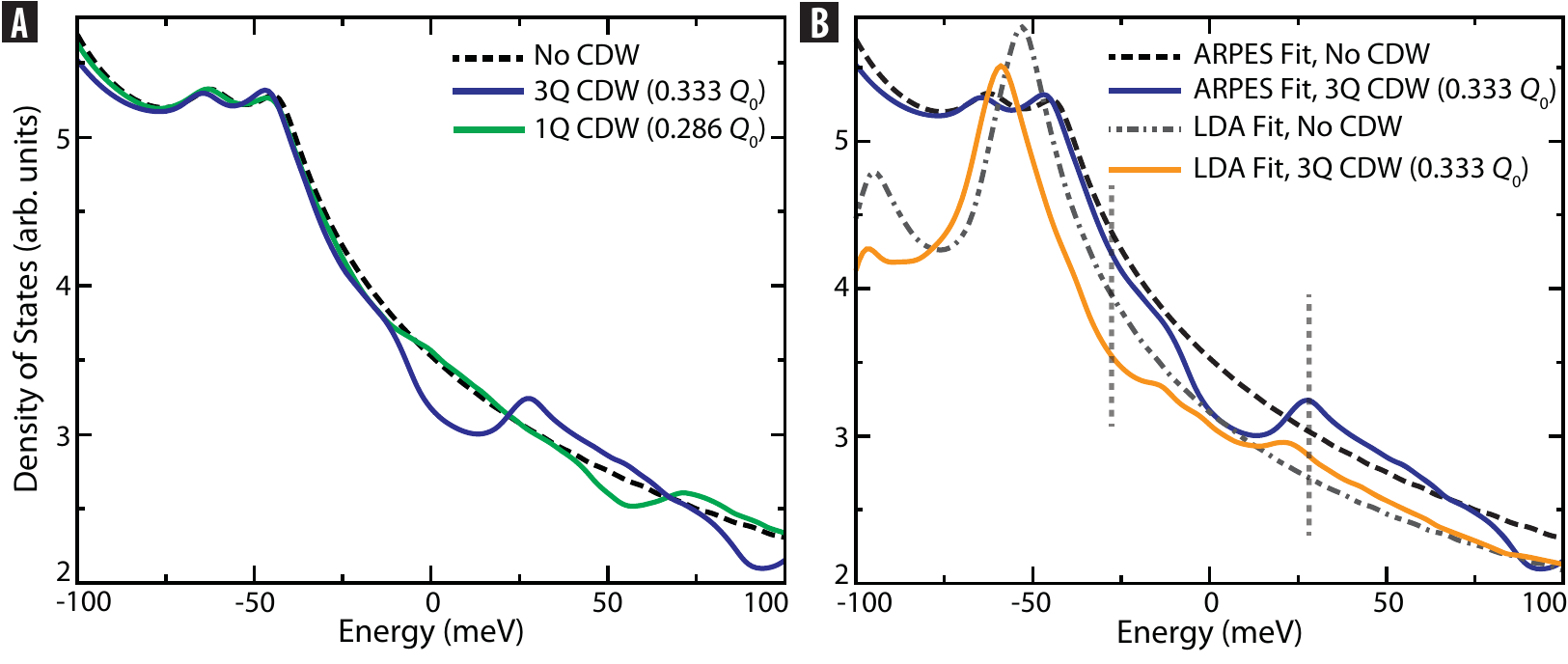}

\caption{\textbf{Comparisons of DOS Calculations.} (\textbf{\emph{A}}) Calculated
DOS spectrum using the ARPES tight-binding fit - in the presence of
a $3Q$ CDW ($\tilde{q}=0.333\, Q_{0}$, blue) and a $1Q$ CDW ($q_{{\rm 1Q}}=0.286\, Q_{0}$,
green), with $\tilde{\Delta}=12\,{\rm meV}$. (\textbf{\emph{B}})
Calculated DOS spectrum using the tight-binding fit to ARPES data\cite{Rahn2012}
compared with a tight-binding fit to LDA calculations\cite{Johannes2006}
in the 'normal' state and in the presence of a $3Q$ CDW ($\tilde{q}=0.333\, Q_{0}$,
$\tilde{\Delta}=12\,{\rm meV}$). The DOS spectrum calculated using
the fit to the ARPES data better reproduces the gap feature observed
in the STM $dI/dV$ spectrum in \ref{fig:spectroscopy}\emph{C}.\label{fig:som_1qldafits}}
\end{figure*}

For completeness, we show in \ref{fig:som_1qldafits}\emph{A} the
calculated DOS in the presence of a $1Q$ CDW at the experimentally
observed wave vector, $q_{{\rm 1Q}}=0.286\, Q_{0}$, using $\tilde{\Delta}=12\,{\rm meV}$.
The lack of correspondence between this calculation (green curve in
\ref{fig:som_1qldafits}\emph{A}) and the measured $dI/dV$ spectrum
(green curve in \ref{fig:spectroscopy}\emph{B}) can be attributed
to the increased intensity of the inelastic background in the buckled
region of the $1Q$ ribbons. This background is evident in the $V$-shape
of the $dI/dV$ spectrum in \ref{fig:spectroscopy}\emph{B}, centered
close to $\varepsilon_{{\rm F}}$, as explained theoretically\cite{Kirtley1992}
and observed experimentally across a wide variety of materials\cite{Kirtley1992,Kirtley1990,Niestemski2007,Fridman2011,Arai2001,Rahnejat2011,Collins1984,McMillan1981}.

We also compare our results to a calculation of the DOS based on a
tight-binding fit to the full three-dimensional LDA band structure
reported by Johannes \emph{et al.}\cite{Johannes2006}. The DOS obtained
using this 3D LDA fit is compared to the results based on the 2D fit
in \ref{fig:som_1qldafits}\emph{B}. We find that the STM data are
best reproduced using the band structure observed by ARPES, and that
there is a noticeable difference between the depths of the gaps in
the two-dimensional ARPES based and three-dimensional LDA-based band
structure fits (despite independent parameter optimization), which
is indicative of some difference between the surface and bulk dispersions\cite{Johannes2006}.

\begin{thebibliography}{50}
\expandafter\ifx\csname natexlab\endcsname\relax\def\natexlab#1{#1}\fi
\expandafter\ifx\csname bibnamefont\endcsname\relax
  \def\bibnamefont#1{#1}\fi
\expandafter\ifx\csname bibfnamefont\endcsname\relax
  \def\bibfnamefont#1{#1}\fi
\expandafter\ifx\csname citenamefont\endcsname\relax
  \def\citenamefont#1{#1}\fi
\expandafter\ifx\csname url\endcsname\relax
  \def\url#1{\texttt{#1}}\fi
\expandafter\ifx\csname urlprefix\endcsname\relax\def\urlprefix{URL }\fi
\providecommand{\bibinfo}[2]{#2}
\providecommand{\eprint}[2][]{\url{#2}}

\bibitem[{\citenamefont{Sachdev}(2000)}]{Sachdev2000}
\bibinfo{author}{\bibfnamefont{S.}~\bibnamefont{Sachdev}},
  \bibinfo{journal}{Science} \textbf{\bibinfo{volume}{288}},
  \bibinfo{pages}{475} (\bibinfo{year}{2000}).

\bibitem[{\citenamefont{Norman et~al.}(2005)\citenamefont{Norman, Pines, and
  Kallin}}]{Norman2005}
\bibinfo{author}{\bibfnamefont{M.~R.} \bibnamefont{Norman}},
  \bibinfo{author}{\bibfnamefont{D.}~\bibnamefont{Pines}}, \bibnamefont{and}
  \bibinfo{author}{\bibfnamefont{C.}~\bibnamefont{Kallin}},
  \bibinfo{journal}{Advances in Physics} \textbf{\bibinfo{volume}{54}},
  \bibinfo{pages}{715} (\bibinfo{year}{2005}).

\bibitem[{\citenamefont{Johnston}(2010)}]{Johnston2010a}
\bibinfo{author}{\bibfnamefont{D.~C.} \bibnamefont{Johnston}},
  \bibinfo{journal}{Advances in Physics} \textbf{\bibinfo{volume}{59}},
  \bibinfo{pages}{803} (\bibinfo{year}{2010}).

\bibitem[{\citenamefont{J\'{e}rome and Schulz}(2002)}]{Jerome2002}
\bibinfo{author}{\bibfnamefont{D.}~\bibnamefont{J\'{e}rome}} \bibnamefont{and}
  \bibinfo{author}{\bibfnamefont{H.~J.} \bibnamefont{Schulz}},
  \bibinfo{journal}{Advances in Physics} \textbf{\bibinfo{volume}{51}},
  \bibinfo{pages}{293} (\bibinfo{year}{2002}).

\bibitem[{\citenamefont{Kivelson et~al.}(2003)\citenamefont{Kivelson, Bindloss,
  Oganesyan, Tranquada, Kapitulnik, and Howald}}]{Kivelson2003}
\bibinfo{author}{\bibfnamefont{S.~A.} \bibnamefont{Kivelson}},
  \bibinfo{author}{\bibfnamefont{I.~P.} \bibnamefont{Bindloss}},
  \bibinfo{author}{\bibfnamefont{V.}~\bibnamefont{Oganesyan}},
  \bibinfo{author}{\bibfnamefont{J.~M.} \bibnamefont{Tranquada}},
  \bibinfo{author}{\bibfnamefont{A.}~\bibnamefont{Kapitulnik}},
  \bibnamefont{and} \bibinfo{author}{\bibfnamefont{C.}~\bibnamefont{Howald}},
  \bibinfo{journal}{Reviews of Modern Physics} \textbf{\bibinfo{volume}{75}},
  \bibinfo{pages}{1201} (\bibinfo{year}{2003}).

\bibitem[{\citenamefont{Johannes et~al.}(2006)\citenamefont{Johannes, Mazin,
  and Howells}}]{Johannes2006}
\bibinfo{author}{\bibfnamefont{M.~D.} \bibnamefont{Johannes}},
  \bibinfo{author}{\bibfnamefont{I.~I.} \bibnamefont{Mazin}}, \bibnamefont{and}
  \bibinfo{author}{\bibfnamefont{C.~A.} \bibnamefont{Howells}},
  \bibinfo{journal}{Physical Review B} \textbf{\bibinfo{volume}{73}},
  \bibinfo{pages}{205102} (\bibinfo{year}{2006}).

\bibitem[{\citenamefont{Suderow et~al.}(2005)\citenamefont{Suderow, Tissen,
  Brison, Mart\'{\i}nez, and Vieira}}]{Suderow2005}
\bibinfo{author}{\bibfnamefont{H.}~\bibnamefont{Suderow}},
  \bibinfo{author}{\bibfnamefont{V.~G.} \bibnamefont{Tissen}},
  \bibinfo{author}{\bibfnamefont{J.~P.} \bibnamefont{Brison}},
  \bibinfo{author}{\bibfnamefont{J.~L.} \bibnamefont{Mart\'{\i}nez}},
  \bibnamefont{and} \bibinfo{author}{\bibfnamefont{S.}~\bibnamefont{Vieira}},
  \bibinfo{journal}{Physical Review Letters} \textbf{\bibinfo{volume}{95}},
  \bibinfo{pages}{117006} (\bibinfo{year}{2005}).

\bibitem[{\citenamefont{Kiss et~al.}(2007)\citenamefont{Kiss, Yokoya, Chainani,
  Shin, Hanaguri, Nohara, and Takagi}}]{Kiss2007}
\bibinfo{author}{\bibfnamefont{T.}~\bibnamefont{Kiss}},
  \bibinfo{author}{\bibfnamefont{T.}~\bibnamefont{Yokoya}},
  \bibinfo{author}{\bibfnamefont{A.}~\bibnamefont{Chainani}},
  \bibinfo{author}{\bibfnamefont{S.}~\bibnamefont{Shin}},
  \bibinfo{author}{\bibfnamefont{T.}~\bibnamefont{Hanaguri}},
  \bibinfo{author}{\bibfnamefont{M.}~\bibnamefont{Nohara}}, \bibnamefont{and}
  \bibinfo{author}{\bibfnamefont{H.}~\bibnamefont{Takagi}},
  \bibinfo{journal}{Nature Physics} \textbf{\bibinfo{volume}{3}},
  \bibinfo{pages}{720} (\bibinfo{year}{2007}).

\bibitem[{\citenamefont{Feng et~al.}(2012)\citenamefont{Feng, Wang, Jaramillo,
  van Wezel, Haravifard, Srajer, Liu, Xu, Littlewood, and
  Rosenbaum}}]{Feng2011}
\bibinfo{author}{\bibfnamefont{Y.}~\bibnamefont{Feng}},
  \bibinfo{author}{\bibfnamefont{J.}~\bibnamefont{Wang}},
  \bibinfo{author}{\bibfnamefont{R.}~\bibnamefont{Jaramillo}},
  \bibinfo{author}{\bibfnamefont{J.}~\bibnamefont{van Wezel}},
  \bibinfo{author}{\bibfnamefont{S.}~\bibnamefont{Haravifard}},
  \bibinfo{author}{\bibfnamefont{G.}~\bibnamefont{Srajer}},
  \bibinfo{author}{\bibfnamefont{Y.}~\bibnamefont{Liu}},
  \bibinfo{author}{\bibfnamefont{Z.~A.} \bibnamefont{Xu}},
  \bibinfo{author}{\bibfnamefont{P.~B.} \bibnamefont{Littlewood}},
  \bibnamefont{and} \bibinfo{author}{\bibfnamefont{T.~F.}
  \bibnamefont{Rosenbaum}}, \bibinfo{journal}{Proceedings of the National
  Academy of Sciences} \textbf{\bibinfo{volume}{109}}, \bibinfo{pages}{7224}
  (\bibinfo{year}{2012}).

\bibitem[{\citenamefont{Moncton et~al.}(1975)\citenamefont{Moncton, Axe, and
  DiSalvo}}]{Moncton1975}
\bibinfo{author}{\bibfnamefont{D.~E.} \bibnamefont{Moncton}},
  \bibinfo{author}{\bibfnamefont{J.~D.} \bibnamefont{Axe}}, \bibnamefont{and}
  \bibinfo{author}{\bibfnamefont{F.~J.} \bibnamefont{DiSalvo}},
  \bibinfo{journal}{Physical Review Letters} \textbf{\bibinfo{volume}{34}},
  \bibinfo{pages}{734} (\bibinfo{year}{1975}).

\bibitem[{\citenamefont{Wilson et~al.}(2001)\citenamefont{Wilson, Salvo, and
  Mahajan}}]{Wilson2001}
\bibinfo{author}{\bibfnamefont{J.~A.} \bibnamefont{Wilson}},
  \bibinfo{author}{\bibfnamefont{F.~J.~D.} \bibnamefont{Salvo}},
  \bibnamefont{and} \bibinfo{author}{\bibfnamefont{S.}~\bibnamefont{Mahajan}},
  \bibinfo{journal}{Advances in Physics} \textbf{\bibinfo{volume}{50}},
  \bibinfo{pages}{1171} (\bibinfo{year}{2001}).

\bibitem[{\citenamefont{Borisenko et~al.}(2009)\citenamefont{Borisenko,
  Kordyuk, Zabolotnyy, Inosov, Evtushinsky, B\"{u}chner, Yaresko, Varykhalov,
  Follath, Eberhardt et~al.}}]{Borisenko2009}
\bibinfo{author}{\bibfnamefont{S.~V.} \bibnamefont{Borisenko}},
  \bibinfo{author}{\bibfnamefont{A.~A.} \bibnamefont{Kordyuk}},
  \bibinfo{author}{\bibfnamefont{V.~B.} \bibnamefont{Zabolotnyy}},
  \bibinfo{author}{\bibfnamefont{D.~S.} \bibnamefont{Inosov}},
  \bibinfo{author}{\bibfnamefont{D.~V.} \bibnamefont{Evtushinsky}},
  \bibinfo{author}{\bibfnamefont{B.}~\bibnamefont{B\"{u}chner}},
  \bibinfo{author}{\bibfnamefont{A.~N.} \bibnamefont{Yaresko}},
  \bibinfo{author}{\bibfnamefont{A.}~\bibnamefont{Varykhalov}},
  \bibinfo{author}{\bibfnamefont{R.}~\bibnamefont{Follath}},
  \bibinfo{author}{\bibfnamefont{W.}~\bibnamefont{Eberhardt}},
  \bibnamefont{et~al.}, \bibinfo{journal}{Physical Review Letters}
  \textbf{\bibinfo{volume}{102}}, \bibinfo{pages}{166402}
  (\bibinfo{year}{2009}).

\bibitem[{\citenamefont{Mialitsin}(2010)}]{Mialitsin2010}
\bibinfo{author}{\bibfnamefont{A.}~\bibnamefont{Mialitsin}},
  \bibinfo{type}{Phd}, \bibinfo{school}{Rutgers University}
  (\bibinfo{year}{2010}).

\bibitem[{\citenamefont{Weber et~al.}(2011)\citenamefont{Weber, Rosenkranz,
  Castellan, Osborn, Hott, Heid, Bohnen, Egami, Said, and Reznik}}]{Weber2011}
\bibinfo{author}{\bibfnamefont{F.}~\bibnamefont{Weber}},
  \bibinfo{author}{\bibfnamefont{S.}~\bibnamefont{Rosenkranz}},
  \bibinfo{author}{\bibfnamefont{J.-P.} \bibnamefont{Castellan}},
  \bibinfo{author}{\bibfnamefont{R.}~\bibnamefont{Osborn}},
  \bibinfo{author}{\bibfnamefont{R.}~\bibnamefont{Hott}},
  \bibinfo{author}{\bibfnamefont{R.}~\bibnamefont{Heid}},
  \bibinfo{author}{\bibfnamefont{K.-P.} \bibnamefont{Bohnen}},
  \bibinfo{author}{\bibfnamefont{T.}~\bibnamefont{Egami}},
  \bibinfo{author}{\bibfnamefont{A.}~\bibnamefont{Said}}, \bibnamefont{and}
  \bibinfo{author}{\bibfnamefont{D.}~\bibnamefont{Reznik}},
  \bibinfo{journal}{Physical Review Letters} \textbf{\bibinfo{volume}{107}},
  \bibinfo{pages}{107403} (\bibinfo{year}{2011}).

\bibitem[{\citenamefont{Rahn et~al.}(2012)\citenamefont{Rahn, Hellmann,
  Kall\"{a}ne, Sohrt, Kim, Kipp, and Rossnagel}}]{Rahn2012}
\bibinfo{author}{\bibfnamefont{D.~J.} \bibnamefont{Rahn}},
  \bibinfo{author}{\bibfnamefont{S.}~\bibnamefont{Hellmann}},
  \bibinfo{author}{\bibfnamefont{M.}~\bibnamefont{Kall\"{a}ne}},
  \bibinfo{author}{\bibfnamefont{C.}~\bibnamefont{Sohrt}},
  \bibinfo{author}{\bibfnamefont{T.~K.} \bibnamefont{Kim}},
  \bibinfo{author}{\bibfnamefont{L.}~\bibnamefont{Kipp}}, \bibnamefont{and}
  \bibinfo{author}{\bibfnamefont{K.}~\bibnamefont{Rossnagel}},
  \bibinfo{journal}{Physical Review B} \textbf{\bibinfo{volume}{85}},
  \bibinfo{pages}{224532} (\bibinfo{year}{2012}).

\bibitem[{\citenamefont{Straub et~al.}(1999)\citenamefont{Straub, Finteis,
  Claessen, Steiner, H\"{u}fner, Blaha, Oglesby, and Bucher}}]{Straub1999}
\bibinfo{author}{\bibfnamefont{T.}~\bibnamefont{Straub}},
  \bibinfo{author}{\bibfnamefont{T.}~\bibnamefont{Finteis}},
  \bibinfo{author}{\bibfnamefont{R.}~\bibnamefont{Claessen}},
  \bibinfo{author}{\bibfnamefont{P.}~\bibnamefont{Steiner}},
  \bibinfo{author}{\bibfnamefont{S.}~\bibnamefont{H\"{u}fner}},
  \bibinfo{author}{\bibfnamefont{P.}~\bibnamefont{Blaha}},
  \bibinfo{author}{\bibfnamefont{C.~S.} \bibnamefont{Oglesby}},
  \bibnamefont{and} \bibinfo{author}{\bibfnamefont{E.}~\bibnamefont{Bucher}},
  \bibinfo{journal}{Physical Review Letters} \textbf{\bibinfo{volume}{82}},
  \bibinfo{pages}{4504} (\bibinfo{year}{1999}).

\bibitem[{\citenamefont{Rossnagel et~al.}(2001)\citenamefont{Rossnagel,
  Seifarth, Kipp, Skibowski, Vo\ss, Kr\"{u}ger, Mazur, and
  Pollmann}}]{Rossnagel2001}
\bibinfo{author}{\bibfnamefont{K.}~\bibnamefont{Rossnagel}},
  \bibinfo{author}{\bibfnamefont{O.}~\bibnamefont{Seifarth}},
  \bibinfo{author}{\bibfnamefont{L.}~\bibnamefont{Kipp}},
  \bibinfo{author}{\bibfnamefont{M.}~\bibnamefont{Skibowski}},
  \bibinfo{author}{\bibfnamefont{D.}~\bibnamefont{Vo\ss}},
  \bibinfo{author}{\bibfnamefont{P.}~\bibnamefont{Kr\"{u}ger}},
  \bibinfo{author}{\bibfnamefont{A.}~\bibnamefont{Mazur}}, \bibnamefont{and}
  \bibinfo{author}{\bibfnamefont{J.}~\bibnamefont{Pollmann}},
  \bibinfo{journal}{Physical Review B} \textbf{\bibinfo{volume}{64}},
  \bibinfo{pages}{235119} (\bibinfo{year}{2001}).

\bibitem[{\citenamefont{Valla et~al.}(2004)\citenamefont{Valla, Fedorov,
  Johnson, Glans, McGuinness, Smith, Andrei, and Berger}}]{Valla2004}
\bibinfo{author}{\bibfnamefont{T.}~\bibnamefont{Valla}},
  \bibinfo{author}{\bibfnamefont{A.~V.} \bibnamefont{Fedorov}},
  \bibinfo{author}{\bibfnamefont{P.~D.} \bibnamefont{Johnson}},
  \bibinfo{author}{\bibfnamefont{P.-A.} \bibnamefont{Glans}},
  \bibinfo{author}{\bibfnamefont{C.}~\bibnamefont{McGuinness}},
  \bibinfo{author}{\bibfnamefont{K.}~\bibnamefont{Smith}},
  \bibinfo{author}{\bibfnamefont{E.~Y.} \bibnamefont{Andrei}},
  \bibnamefont{and} \bibinfo{author}{\bibfnamefont{H.}~\bibnamefont{Berger}},
  \bibinfo{journal}{Physical Review Letters} \textbf{\bibinfo{volume}{92}},
  \bibinfo{pages}{086401} (\bibinfo{year}{2004}).

\bibitem[{\citenamefont{Shen et~al.}(2008)\citenamefont{Shen, Zhang, Yang, Wei,
  Ou, Dong, Xie, He, Zhao, Zhou et~al.}}]{Shen2008a}
\bibinfo{author}{\bibfnamefont{D.~W.} \bibnamefont{Shen}},
  \bibinfo{author}{\bibfnamefont{Y.}~\bibnamefont{Zhang}},
  \bibinfo{author}{\bibfnamefont{L.~X.} \bibnamefont{Yang}},
  \bibinfo{author}{\bibfnamefont{J.}~\bibnamefont{Wei}},
  \bibinfo{author}{\bibfnamefont{H.~W.} \bibnamefont{Ou}},
  \bibinfo{author}{\bibfnamefont{J.~K.} \bibnamefont{Dong}},
  \bibinfo{author}{\bibfnamefont{B.~P.} \bibnamefont{Xie}},
  \bibinfo{author}{\bibfnamefont{C.}~\bibnamefont{He}},
  \bibinfo{author}{\bibfnamefont{J.~F.} \bibnamefont{Zhao}},
  \bibinfo{author}{\bibfnamefont{B.}~\bibnamefont{Zhou}}, \bibnamefont{et~al.},
  \bibinfo{journal}{Physical Review Letters} \textbf{\bibinfo{volume}{101}},
  \bibinfo{pages}{226406} (\bibinfo{year}{2008}).

\bibitem[{\citenamefont{Hess et~al.}(1991)\citenamefont{Hess, Robinson, and
  Waszczak}}]{Hess1991}
\bibinfo{author}{\bibfnamefont{H.~F.} \bibnamefont{Hess}},
  \bibinfo{author}{\bibfnamefont{R.~B.} \bibnamefont{Robinson}},
  \bibnamefont{and} \bibinfo{author}{\bibfnamefont{J.~V.}
  \bibnamefont{Waszczak}}, \bibinfo{journal}{Physica B: Condensed Matter}
  \textbf{\bibinfo{volume}{169}}, \bibinfo{pages}{422} (\bibinfo{year}{1991}).

\bibitem[{\citenamefont{McMillan}(1976)}]{McMillan1976}
\bibinfo{author}{\bibfnamefont{W.~L.} \bibnamefont{McMillan}},
  \bibinfo{journal}{Physical Review B} \textbf{\bibinfo{volume}{14}},
  \bibinfo{pages}{1496} (\bibinfo{year}{1976}).

\bibitem[{\citenamefont{Wang et~al.}(2009)\citenamefont{Wang, Lee, Dreyer, and
  Barker}}]{Wang2009g}
\bibinfo{author}{\bibfnamefont{H.}~\bibnamefont{Wang}},
  \bibinfo{author}{\bibfnamefont{J.}~\bibnamefont{Lee}},
  \bibinfo{author}{\bibfnamefont{M.}~\bibnamefont{Dreyer}}, \bibnamefont{and}
  \bibinfo{author}{\bibfnamefont{B.~I.} \bibnamefont{Barker}},
  \bibinfo{journal}{Journal of Physics: Condensed Matter}
  \textbf{\bibinfo{volume}{21}}, \bibinfo{pages}{265005}
  (\bibinfo{year}{2009}), ISSN \bibinfo{issn}{0953-8984}.

\bibitem[{\citenamefont{Bando et~al.}(1997)\citenamefont{Bando, Miyahara,
  Enomoto, and Ozaki}}]{Bando1997}
\bibinfo{author}{\bibfnamefont{H.}~\bibnamefont{Bando}},
  \bibinfo{author}{\bibfnamefont{Y.}~\bibnamefont{Miyahara}},
  \bibinfo{author}{\bibfnamefont{H.}~\bibnamefont{Enomoto}}, \bibnamefont{and}
  \bibinfo{author}{\bibfnamefont{H.}~\bibnamefont{Ozaki}},
  \bibinfo{journal}{Surface Science} \textbf{\bibinfo{volume}{381}},
  \bibinfo{pages}{L609} (\bibinfo{year}{1997}).

\bibitem[{\citenamefont{McMillan}(1975)}]{McMillan1975}
\bibinfo{author}{\bibfnamefont{W.~L.} \bibnamefont{McMillan}},
  \bibinfo{journal}{Physical Review B} \textbf{\bibinfo{volume}{12}},
  \bibinfo{pages}{1187} (\bibinfo{year}{1975}).

\bibitem[{\citenamefont{Norman et~al.}(2007)\citenamefont{Norman, Kanigel,
  Randeria, Chatterjee, and Campuzano}}]{Norman2007}
\bibinfo{author}{\bibfnamefont{M.~R.} \bibnamefont{Norman}},
  \bibinfo{author}{\bibfnamefont{A.}~\bibnamefont{Kanigel}},
  \bibinfo{author}{\bibfnamefont{M.}~\bibnamefont{Randeria}},
  \bibinfo{author}{\bibfnamefont{U.}~\bibnamefont{Chatterjee}},
  \bibnamefont{and} \bibinfo{author}{\bibfnamefont{J.~C.}
  \bibnamefont{Campuzano}}, \bibinfo{journal}{Physical Review B}
  \textbf{\bibinfo{volume}{76}}, \bibinfo{pages}{174501}
  (\bibinfo{year}{2007}).

\bibitem[{\citenamefont{Kirtley et~al.}(1992)\citenamefont{Kirtley, Washburn,
  and Scalapino}}]{Kirtley1992}
\bibinfo{author}{\bibfnamefont{J.~R.} \bibnamefont{Kirtley}},
  \bibinfo{author}{\bibfnamefont{S.}~\bibnamefont{Washburn}}, \bibnamefont{and}
  \bibinfo{author}{\bibfnamefont{D.~J.} \bibnamefont{Scalapino}},
  \bibinfo{journal}{Physical Review B} \textbf{\bibinfo{volume}{45}},
  \bibinfo{pages}{336} (\bibinfo{year}{1992}).

\bibitem[{\citenamefont{Yao et~al.}(2006)\citenamefont{Yao, Robertson, Kim, and
  Kivelson}}]{Yao2006}
\bibinfo{author}{\bibfnamefont{H.}~\bibnamefont{Yao}},
  \bibinfo{author}{\bibfnamefont{J.~A.} \bibnamefont{Robertson}},
  \bibinfo{author}{\bibfnamefont{E.-A.} \bibnamefont{Kim}}, \bibnamefont{and}
  \bibinfo{author}{\bibfnamefont{S.~A.} \bibnamefont{Kivelson}},
  \bibinfo{journal}{Physical Review B} \textbf{\bibinfo{volume}{74}},
  \bibinfo{pages}{245126} (\bibinfo{year}{2006}).

\bibitem[{\citenamefont{Ru et~al.}(2008)\citenamefont{Ru, Condron, Margulis,
  Shin, Laverock, Dugdale, Toney, and Fisher}}]{Ru2008a}
\bibinfo{author}{\bibfnamefont{N.}~\bibnamefont{Ru}},
  \bibinfo{author}{\bibfnamefont{C.~L.} \bibnamefont{Condron}},
  \bibinfo{author}{\bibfnamefont{G.~Y.} \bibnamefont{Margulis}},
  \bibinfo{author}{\bibfnamefont{K.~Y.} \bibnamefont{Shin}},
  \bibinfo{author}{\bibfnamefont{J.}~\bibnamefont{Laverock}},
  \bibinfo{author}{\bibfnamefont{S.~B.} \bibnamefont{Dugdale}},
  \bibinfo{author}{\bibfnamefont{M.~F.} \bibnamefont{Toney}}, \bibnamefont{and}
  \bibinfo{author}{\bibfnamefont{I.~R.} \bibnamefont{Fisher}},
  \bibinfo{journal}{Physical Review B} \textbf{\bibinfo{volume}{77}},
  \bibinfo{pages}{035114} (\bibinfo{year}{2008}).

\bibitem[{\citenamefont{Slezak et~al.}(2008)\citenamefont{Slezak, Lee, Wang,
  McElroy, Fujita, Andersen, Hirschfeld, Eisaki, Uchida, and
  Davis}}]{Slezak2008}
\bibinfo{author}{\bibfnamefont{J.~A.} \bibnamefont{Slezak}},
  \bibinfo{author}{\bibfnamefont{J.}~\bibnamefont{Lee}},
  \bibinfo{author}{\bibfnamefont{M.}~\bibnamefont{Wang}},
  \bibinfo{author}{\bibfnamefont{K.}~\bibnamefont{McElroy}},
  \bibinfo{author}{\bibfnamefont{K.}~\bibnamefont{Fujita}},
  \bibinfo{author}{\bibfnamefont{B.~M.} \bibnamefont{Andersen}},
  \bibinfo{author}{\bibfnamefont{P.~J.} \bibnamefont{Hirschfeld}},
  \bibinfo{author}{\bibfnamefont{H.}~\bibnamefont{Eisaki}},
  \bibinfo{author}{\bibfnamefont{S.}~\bibnamefont{Uchida}}, \bibnamefont{and}
  \bibinfo{author}{\bibfnamefont{J.~C.} \bibnamefont{Davis}},
  \bibinfo{journal}{Proceedings of the National Academy of Sciences}
  \textbf{\bibinfo{volume}{105}}, \bibinfo{pages}{3203} (\bibinfo{year}{2008}).

\bibitem[{\citenamefont{Chu et~al.}(2010)\citenamefont{Chu, Analytis, {De
  Greve}, McMahon, Islam, Yamamoto, and Fisher}}]{Chu2010a}
\bibinfo{author}{\bibfnamefont{J.-H.} \bibnamefont{Chu}},
  \bibinfo{author}{\bibfnamefont{J.~G.} \bibnamefont{Analytis}},
  \bibinfo{author}{\bibfnamefont{K.}~\bibnamefont{{De Greve}}},
  \bibinfo{author}{\bibfnamefont{P.~L.} \bibnamefont{McMahon}},
  \bibinfo{author}{\bibfnamefont{Z.}~\bibnamefont{Islam}},
  \bibinfo{author}{\bibfnamefont{Y.}~\bibnamefont{Yamamoto}}, \bibnamefont{and}
  \bibinfo{author}{\bibfnamefont{I.~R.} \bibnamefont{Fisher}},
  \bibinfo{journal}{Science} \textbf{\bibinfo{volume}{329}},
  \bibinfo{pages}{824} (\bibinfo{year}{2010}).

\bibitem[{\citenamefont{Robertson et~al.}(2006)\citenamefont{Robertson,
  Kivelson, Fradkin, Fang, and Kapitulnik}}]{Robertson2006}
\bibinfo{author}{\bibfnamefont{J.~A.} \bibnamefont{Robertson}},
  \bibinfo{author}{\bibfnamefont{S.~A.} \bibnamefont{Kivelson}},
  \bibinfo{author}{\bibfnamefont{E.}~\bibnamefont{Fradkin}},
  \bibinfo{author}{\bibfnamefont{A.~C.} \bibnamefont{Fang}}, \bibnamefont{and}
  \bibinfo{author}{\bibfnamefont{A.}~\bibnamefont{Kapitulnik}},
  \bibinfo{journal}{Physical Review B} \textbf{\bibinfo{volume}{74}},
  \bibinfo{pages}{134507} (\bibinfo{year}{2006}).

\bibitem[{\citenamefont{{Del Maestro} et~al.}(2006)\citenamefont{{Del Maestro},
  Rosenow, and Sachdev}}]{DelMaestro2006a}
\bibinfo{author}{\bibfnamefont{A.}~\bibnamefont{{Del Maestro}}},
  \bibinfo{author}{\bibfnamefont{B.}~\bibnamefont{Rosenow}}, \bibnamefont{and}
  \bibinfo{author}{\bibfnamefont{S.}~\bibnamefont{Sachdev}},
  \bibinfo{journal}{Physical Review B} \textbf{\bibinfo{volume}{74}},
  \bibinfo{pages}{024520} (\bibinfo{year}{2006}).

\bibitem[{\citenamefont{Wise et~al.}(2008)\citenamefont{Wise, Boyer,
  Chatterjee, Kondo, Takeuchi, Ikuta, Wang, and Hudson}}]{Wise2008}
\bibinfo{author}{\bibfnamefont{W.~D.} \bibnamefont{Wise}},
  \bibinfo{author}{\bibfnamefont{M.~C.} \bibnamefont{Boyer}},
  \bibinfo{author}{\bibfnamefont{K.}~\bibnamefont{Chatterjee}},
  \bibinfo{author}{\bibfnamefont{T.}~\bibnamefont{Kondo}},
  \bibinfo{author}{\bibfnamefont{T.}~\bibnamefont{Takeuchi}},
  \bibinfo{author}{\bibfnamefont{H.}~\bibnamefont{Ikuta}},
  \bibinfo{author}{\bibfnamefont{Y.}~\bibnamefont{Wang}}, \bibnamefont{and}
  \bibinfo{author}{\bibfnamefont{E.~W.} \bibnamefont{Hudson}},
  \bibinfo{journal}{Nature Physics} \textbf{\bibinfo{volume}{4}},
  \bibinfo{pages}{696} (\bibinfo{year}{2008}).

\bibitem[{\citenamefont{Parker et~al.}(2010)\citenamefont{Parker, Aynajian, {da
  Silva Neto}, Pushp, Ono, Wen, Xu, Gu, and Yazdani}}]{Parker2010}
\bibinfo{author}{\bibfnamefont{C.~V.} \bibnamefont{Parker}},
  \bibinfo{author}{\bibfnamefont{P.}~\bibnamefont{Aynajian}},
  \bibinfo{author}{\bibfnamefont{E.~H.} \bibnamefont{{da Silva Neto}}},
  \bibinfo{author}{\bibfnamefont{A.}~\bibnamefont{Pushp}},
  \bibinfo{author}{\bibfnamefont{S.}~\bibnamefont{Ono}},
  \bibinfo{author}{\bibfnamefont{J.}~\bibnamefont{Wen}},
  \bibinfo{author}{\bibfnamefont{Z.}~\bibnamefont{Xu}},
  \bibinfo{author}{\bibfnamefont{G.}~\bibnamefont{Gu}}, \bibnamefont{and}
  \bibinfo{author}{\bibfnamefont{A.}~\bibnamefont{Yazdani}},
  \bibinfo{journal}{Nature} \textbf{\bibinfo{volume}{468}},
  \bibinfo{pages}{677} (\bibinfo{year}{2010}).

\bibitem[{\citenamefont{Zeljkovic et~al.}(2012)\citenamefont{Zeljkovic, Xu,
  Wen, Gu, Markiewicz, and Hoffman}}]{Zeljkovic2012b}
\bibinfo{author}{\bibfnamefont{I.}~\bibnamefont{Zeljkovic}},
  \bibinfo{author}{\bibfnamefont{Z.}~\bibnamefont{Xu}},
  \bibinfo{author}{\bibfnamefont{J.}~\bibnamefont{Wen}},
  \bibinfo{author}{\bibfnamefont{G.}~\bibnamefont{Gu}},
  \bibinfo{author}{\bibfnamefont{R.~S.} \bibnamefont{Markiewicz}},
  \bibnamefont{and} \bibinfo{author}{\bibfnamefont{J.~E.}
  \bibnamefont{Hoffman}}, \bibinfo{journal}{Science}
  \textbf{\bibinfo{volume}{337}}, \bibinfo{pages}{320} (\bibinfo{year}{2012}).

\bibitem[{\citenamefont{Saha et~al.}(2012)\citenamefont{Saha, Butch, Drye,
  Magill, Ziemak, Kirshenbaum, Zavalij, Lynn, and Paglione}}]{Saha2012}
\bibinfo{author}{\bibfnamefont{S.~R.} \bibnamefont{Saha}},
  \bibinfo{author}{\bibfnamefont{N.~P.} \bibnamefont{Butch}},
  \bibinfo{author}{\bibfnamefont{T.}~\bibnamefont{Drye}},
  \bibinfo{author}{\bibfnamefont{J.}~\bibnamefont{Magill}},
  \bibinfo{author}{\bibfnamefont{S.}~\bibnamefont{Ziemak}},
  \bibinfo{author}{\bibfnamefont{K.}~\bibnamefont{Kirshenbaum}},
  \bibinfo{author}{\bibfnamefont{P.~Y.} \bibnamefont{Zavalij}},
  \bibinfo{author}{\bibfnamefont{J.~W.} \bibnamefont{Lynn}}, \bibnamefont{and}
  \bibinfo{author}{\bibfnamefont{J.}~\bibnamefont{Paglione}},
  \bibinfo{journal}{Physical Review B} \textbf{\bibinfo{volume}{85}},
  \bibinfo{pages}{024525} (\bibinfo{year}{2012}), ISSN
  \bibinfo{issn}{1098-0121}.

\bibitem[{\citenamefont{Wang et~al.}(2012)\citenamefont{Wang, Li, Zhang, Zhang,
  Zhang, Li, Ding, Ou, Deng, Chang et~al.}}]{Wang2012b}
\bibinfo{author}{\bibfnamefont{Q.-Y.} \bibnamefont{Wang}},
  \bibinfo{author}{\bibfnamefont{Z.}~\bibnamefont{Li}},
  \bibinfo{author}{\bibfnamefont{W.-H.} \bibnamefont{Zhang}},
  \bibinfo{author}{\bibfnamefont{Z.-C.} \bibnamefont{Zhang}},
  \bibinfo{author}{\bibfnamefont{J.-S.} \bibnamefont{Zhang}},
  \bibinfo{author}{\bibfnamefont{W.}~\bibnamefont{Li}},
  \bibinfo{author}{\bibfnamefont{H.}~\bibnamefont{Ding}},
  \bibinfo{author}{\bibfnamefont{Y.-B.} \bibnamefont{Ou}},
  \bibinfo{author}{\bibfnamefont{P.}~\bibnamefont{Deng}},
  \bibinfo{author}{\bibfnamefont{K.}~\bibnamefont{Chang}},
  \bibnamefont{et~al.}, \bibinfo{journal}{Chinese Physics Letters}
  \textbf{\bibinfo{volume}{29}}, \bibinfo{pages}{037402}
  (\bibinfo{year}{2012}).

\bibitem[{\citenamefont{Hanaguri et~al.}(2010)\citenamefont{Hanaguri, Igarashi,
  Kawamura, Takagi, and Sasagawa}}]{Hanaguri2010}
\bibinfo{author}{\bibfnamefont{T.}~\bibnamefont{Hanaguri}},
  \bibinfo{author}{\bibfnamefont{K.}~\bibnamefont{Igarashi}},
  \bibinfo{author}{\bibfnamefont{M.}~\bibnamefont{Kawamura}},
  \bibinfo{author}{\bibfnamefont{H.}~\bibnamefont{Takagi}}, \bibnamefont{and}
  \bibinfo{author}{\bibfnamefont{T.}~\bibnamefont{Sasagawa}},
  \bibinfo{journal}{Physical Review B} \textbf{\bibinfo{volume}{82}},
  \bibinfo{pages}{1} (\bibinfo{year}{2010}).

\bibitem[{\citenamefont{Rutter et~al.}(2007)\citenamefont{Rutter, Crain,
  Guisinger, Li, First, and Stroscio}}]{Rutter2007a}
\bibinfo{author}{\bibfnamefont{G.~M.} \bibnamefont{Rutter}},
  \bibinfo{author}{\bibfnamefont{J.~N.} \bibnamefont{Crain}},
  \bibinfo{author}{\bibfnamefont{N.~P.} \bibnamefont{Guisinger}},
  \bibinfo{author}{\bibfnamefont{T.}~\bibnamefont{Li}},
  \bibinfo{author}{\bibfnamefont{P.~N.} \bibnamefont{First}}, \bibnamefont{and}
  \bibinfo{author}{\bibfnamefont{J.~A.} \bibnamefont{Stroscio}},
  \bibinfo{journal}{Science} \textbf{\bibinfo{volume}{317}},
  \bibinfo{pages}{219} (\bibinfo{year}{2007}).

\bibitem[{\citenamefont{Okada et~al.}(2012)\citenamefont{Okada, Zhou, Walkup,
  Dhital, Wilson, and Madhavan}}]{Okada2012c}
\bibinfo{author}{\bibfnamefont{Y.}~\bibnamefont{Okada}},
  \bibinfo{author}{\bibfnamefont{W.}~\bibnamefont{Zhou}},
  \bibinfo{author}{\bibfnamefont{D.}~\bibnamefont{Walkup}},
  \bibinfo{author}{\bibfnamefont{C.}~\bibnamefont{Dhital}},
  \bibinfo{author}{\bibfnamefont{S.~D.} \bibnamefont{Wilson}},
  \bibnamefont{and} \bibinfo{author}{\bibfnamefont{V.}~\bibnamefont{Madhavan}},
  \bibinfo{journal}{Nature Communications} \textbf{\bibinfo{volume}{3}},
  \bibinfo{pages}{1158} (\bibinfo{year}{2012}).

\bibitem[{\citenamefont{Lawler et~al.}(2010)\citenamefont{Lawler, Fujita, Lee,
  Schmidt, Kohsaka, Kim, Eisaki, Uchida, Davis, Sethna et~al.}}]{Lawler2010}
\bibinfo{author}{\bibfnamefont{M.~J.} \bibnamefont{Lawler}},
  \bibinfo{author}{\bibfnamefont{K.}~\bibnamefont{Fujita}},
  \bibinfo{author}{\bibfnamefont{J.}~\bibnamefont{Lee}},
  \bibinfo{author}{\bibfnamefont{A.~R.} \bibnamefont{Schmidt}},
  \bibinfo{author}{\bibfnamefont{Y.}~\bibnamefont{Kohsaka}},
  \bibinfo{author}{\bibfnamefont{C.~K.} \bibnamefont{Kim}},
  \bibinfo{author}{\bibfnamefont{H.}~\bibnamefont{Eisaki}},
  \bibinfo{author}{\bibfnamefont{S.}~\bibnamefont{Uchida}},
  \bibinfo{author}{\bibfnamefont{J.~C.} \bibnamefont{Davis}},
  \bibinfo{author}{\bibfnamefont{J.~P.} \bibnamefont{Sethna}},
  \bibnamefont{et~al.}, \bibinfo{journal}{Nature}
  \textbf{\bibinfo{volume}{466}}, \bibinfo{pages}{347} (\bibinfo{year}{2010}).

\bibitem[{\citenamefont{Hamidian et~al.}(2012)\citenamefont{Hamidian, Firmo,
  Fujita, Mukhopadhyay, Orenstein, Eisaki, Uchida, Lawler, Kim, and
  Davis}}]{Hamidian2012}
\bibinfo{author}{\bibfnamefont{M.~H.} \bibnamefont{Hamidian}},
  \bibinfo{author}{\bibfnamefont{I.~A.} \bibnamefont{Firmo}},
  \bibinfo{author}{\bibfnamefont{K.}~\bibnamefont{Fujita}},
  \bibinfo{author}{\bibfnamefont{S.}~\bibnamefont{Mukhopadhyay}},
  \bibinfo{author}{\bibfnamefont{J.~W.} \bibnamefont{Orenstein}},
  \bibinfo{author}{\bibfnamefont{H.}~\bibnamefont{Eisaki}},
  \bibinfo{author}{\bibfnamefont{S.}~\bibnamefont{Uchida}},
  \bibinfo{author}{\bibfnamefont{M.~J.} \bibnamefont{Lawler}},
  \bibinfo{author}{\bibfnamefont{E.-A.} \bibnamefont{Kim}}, \bibnamefont{and}
  \bibinfo{author}{\bibfnamefont{J.~C.} \bibnamefont{Davis}},
  \bibinfo{journal}{New Journal of Physics} \textbf{\bibinfo{volume}{14}},
  \bibinfo{pages}{053017} (\bibinfo{year}{2012}).

\bibitem[{\citenamefont{Williams}(2011)}]{Williams2011a}
\bibinfo{author}{\bibfnamefont{T.~L.} \bibnamefont{Williams}},
  \bibinfo{type}{Phd}, \bibinfo{school}{Harvard University}
  (\bibinfo{year}{2011}).

\bibitem[{\citenamefont{Kirtley and Scalapino}(1990)}]{Kirtley1990}
\bibinfo{author}{\bibfnamefont{J.~R.} \bibnamefont{Kirtley}} \bibnamefont{and}
  \bibinfo{author}{\bibfnamefont{D.~J.} \bibnamefont{Scalapino}},
  \bibinfo{journal}{Physical Review Letters} \textbf{\bibinfo{volume}{65}},
  \bibinfo{pages}{798} (\bibinfo{year}{1990}).

\bibitem[{\citenamefont{Niestemski et~al.}(2007)\citenamefont{Niestemski,
  Kunwar, Zhou, Li, Ding, Wang, Dai, and Madhavan}}]{Niestemski2007}
\bibinfo{author}{\bibfnamefont{F.~C.} \bibnamefont{Niestemski}},
  \bibinfo{author}{\bibfnamefont{S.}~\bibnamefont{Kunwar}},
  \bibinfo{author}{\bibfnamefont{S.}~\bibnamefont{Zhou}},
  \bibinfo{author}{\bibfnamefont{S.}~\bibnamefont{Li}},
  \bibinfo{author}{\bibfnamefont{H.}~\bibnamefont{Ding}},
  \bibinfo{author}{\bibfnamefont{Z.}~\bibnamefont{Wang}},
  \bibinfo{author}{\bibfnamefont{P.}~\bibnamefont{Dai}}, \bibnamefont{and}
  \bibinfo{author}{\bibfnamefont{V.}~\bibnamefont{Madhavan}},
  \bibinfo{journal}{Nature} \textbf{\bibinfo{volume}{450}},
  \bibinfo{pages}{1058} (\bibinfo{year}{2007}), ISSN \bibinfo{issn}{1476-4687}.

\bibitem[{\citenamefont{Fridman et~al.}(2011)\citenamefont{Fridman, Yeh, Wu,
  and Wei}}]{Fridman2011}
\bibinfo{author}{\bibfnamefont{I.}~\bibnamefont{Fridman}},
  \bibinfo{author}{\bibfnamefont{K.-W.} \bibnamefont{Yeh}},
  \bibinfo{author}{\bibfnamefont{M.-K.} \bibnamefont{Wu}}, \bibnamefont{and}
  \bibinfo{author}{\bibfnamefont{J.~Y.~T.} \bibnamefont{Wei}},
  \bibinfo{journal}{Journal of Physics and Chemistry of Solids}
  \textbf{\bibinfo{volume}{72}}, \bibinfo{pages}{483} (\bibinfo{year}{2011}).

\bibitem[{\citenamefont{Arai et~al.}(2001)\citenamefont{Arai, Ichimura, Nomura,
  Takasaki, Yamada, Nakatsuji, and Anzai}}]{Arai2001}
\bibinfo{author}{\bibfnamefont{T.}~\bibnamefont{Arai}},
  \bibinfo{author}{\bibfnamefont{K.}~\bibnamefont{Ichimura}},
  \bibinfo{author}{\bibfnamefont{K.}~\bibnamefont{Nomura}},
  \bibinfo{author}{\bibfnamefont{S.}~\bibnamefont{Takasaki}},
  \bibinfo{author}{\bibfnamefont{J.}~\bibnamefont{Yamada}},
  \bibinfo{author}{\bibfnamefont{S.}~\bibnamefont{Nakatsuji}},
  \bibnamefont{and} \bibinfo{author}{\bibfnamefont{H.}~\bibnamefont{Anzai}},
  \bibinfo{journal}{Physical Review B} \textbf{\bibinfo{volume}{63}},
  \bibinfo{pages}{104518} (\bibinfo{year}{2001}).

\bibitem[{\citenamefont{Rahnejat et~al.}(2011)\citenamefont{Rahnejat, Howard,
  Shuttleworth, Schofield, Iwaya, Hirjibehedin, Renner, Aeppli, and
  Ellerby}}]{Rahnejat2011}
\bibinfo{author}{\bibfnamefont{K.}~\bibnamefont{Rahnejat}},
  \bibinfo{author}{\bibfnamefont{C.}~\bibnamefont{Howard}},
  \bibinfo{author}{\bibfnamefont{N.}~\bibnamefont{Shuttleworth}},
  \bibinfo{author}{\bibfnamefont{S.}~\bibnamefont{Schofield}},
  \bibinfo{author}{\bibfnamefont{K.}~\bibnamefont{Iwaya}},
  \bibinfo{author}{\bibfnamefont{C.}~\bibnamefont{Hirjibehedin}},
  \bibinfo{author}{\bibfnamefont{C.}~\bibnamefont{Renner}},
  \bibinfo{author}{\bibfnamefont{G.}~\bibnamefont{Aeppli}}, \bibnamefont{and}
  \bibinfo{author}{\bibfnamefont{M.}~\bibnamefont{Ellerby}},
  \bibinfo{journal}{Nature Communications} \textbf{\bibinfo{volume}{2}},
  \bibinfo{pages}{558} (\bibinfo{year}{2011}).

\bibitem[{\citenamefont{Collins et~al.}(1984)\citenamefont{Collins, Lambe,
  McGill, and Burnham}}]{Collins1984}
\bibinfo{author}{\bibfnamefont{R.~T.} \bibnamefont{Collins}},
  \bibinfo{author}{\bibfnamefont{J.}~\bibnamefont{Lambe}},
  \bibinfo{author}{\bibfnamefont{T.~C.} \bibnamefont{McGill}},
  \bibnamefont{and} \bibinfo{author}{\bibfnamefont{R.~D.}
  \bibnamefont{Burnham}}, \bibinfo{journal}{Applied Physics Letters}
  \textbf{\bibinfo{volume}{44}}, \bibinfo{pages}{532} (\bibinfo{year}{1984}).

\bibitem[{\citenamefont{McMillan and Mochel}(1981)}]{McMillan1981}
\bibinfo{author}{\bibfnamefont{W.}~\bibnamefont{McMillan}} \bibnamefont{and}
  \bibinfo{author}{\bibfnamefont{J.}~\bibnamefont{Mochel}},
  \bibinfo{journal}{Physical Review Letters} \textbf{\bibinfo{volume}{46}},
  \bibinfo{pages}{556} (\bibinfo{year}{1981}).

\end{thebibliography}
\end{document}